\documentclass[12pt]{amsart}
\usepackage{palatino}
\usepackage{amscd,amssymb}
\usepackage{float}
\usepackage{graphicx}

\usepackage[cmtip,matrix,arrow]{xy}

\numberwithin{equation}{section}

\newtheorem {Theorem}                   {Theorem}
\newtheorem {varTheorem}                {Theorem}

\newtheorem {Lemma}[equation]           {Lemma}
\newtheorem {Proposition}[equation]     {Proposition}

\newtheorem {Question}[equation]      {Question}

\newtheorem {corollary}[equation]       {Corollary}

\theoremstyle{definition}
\newtheorem {Definition}[equation]{Definition}
\newtheorem{zremark}[equation] {Remark}
\newenvironment{Remark} {\par\footnotesize\zremark}{~\\}{\endzremark}

\newcommand{\pr} {\smallskip\noindent{\bf Proof\,\,}}

\newenvironment{pf}  {\begin{proof}}{\end{proof}}

\newcommand     {\comment}[1]   {}
\newcommand     {\mute}[2] {}
\newcommand     {\printname}[1] {}

\newcommand{\labell}[1] {\label{#1}\printname{#1}}

\def    \to     {\longrightarrow}

\def    \C      {{\mathbb C}}
\def    \R      {{\mathbb R}}

\def    \Z      {{\mathbb Z}}

\def    \cD     {{\mathcal D}}
\def    \cO     {{\mathcal O}}
\def\cB{{\bf B}}
\def\cH{{\bf H}}
\def    \cF     {{\mathcal F}}

\def    \cJ     {{\mathcal J}}
\def    \cR     {{\mathcal R}}

\def    \cA     {{\mathcal A}}
\def    \cbA     {{\bf{ A}}}

\def    \cU    {{\mathcal U}}\def    \cN    {{\mathcal N}}

\def    \fg     {{\mathfrak g}}

\def    \pr     {\operatorname{pr}}

\begin{document}

\title[Reflection Positivity and Chern-Simons Functional Integrals]{Reflection Positivity and Chern-Simons Functional Integrals}

\author{Jonathan Weitsman}
\thanks{Supported in part by a grant from the Simons Foundation (\# 579801 to J.W.)}
\address{Department of Mathematics, Northeastern University, Boston, MA 02115}
\email{j.weitsman@neu.edu}
\thanks{\today}

\begin{abstract} 
We show that a mathematical version of the formal Chern-Simons functional integral of \cite{Witten} for manifolds equipped with a reflection may be  constructed in terms of a reflection positive functional, associated to the quadratic term in the Chern-Simons Lagrangian, on an algebra of functions on a Banach space $\cbA$ of connections on the underlying 3-manifold. This construction yields a Hilbert space associated to a surface preserved by the reflection.  A version of
the cubic Bosonic interaction term in the Chern-Simons Lagrangian gives a self-adjoint operator on this Hilbert space, and by exponentiation, a unitary one parameter subgroup of operators.  The vacuum
expectation value of this one parameter subgroup is combined with an additional term associated to the ghost fields and their interaction, and an appropriate weak limit gives a partition function for the quantum field theory. This construction is nonperturbative.  The theory is finite and does not require renormalization, as may be expected from perturbation theory \cite{as}. It is natural to ask whether the resulting partition function is related to the manifold invariants of Witten and Reshetikhin-Turaev, or whether a more elaborate construction may be needed.\end{abstract}

\maketitle

\tableofcontents

\section{Introduction}
Let $M$ be a compact oriented 3-manifold, let $G$ be a compact Lie group, and let $k \in \Z.$   In \cite{Witten} E. Witten showed that a formal functional integral 
\begin{equation}\labell{ffi}
Z_k(M) = \int_\cA dA e^{i k {\rm CS}(A)}
\end{equation}
\noindent of the exponential of the Chern-Simons functional over the space $\cA$ of connections on the trivial principal $G$-bundle on $M$ should be interpreted as yielding a topological invariant of the three manifold $M,$ and that integrals of certain functions in this formal functional integral should be interpreted as link invariants, which in the case $M = S^3$ and $G = SU(n)$ are the Jones polynomials \cite{Jones} of links.

Invariants satisfying the properties expected from Witten's results were constructed by Reshetikhin and Turaev \cite{rt} and a vast literature has grown in this field which provides evidence that properties expected from the formal functional integral and its formal perturbation expansion are satisfied by these invariants. 

The purpose of this paper is to give a mathematically well-defined construction of a version of the functional integral (\ref{ffi}), for manifolds $M$ equipped with a reflection, in terms of a reflection positive functional on an algebra of functions on an appropriate space of connections.  As in \cite{w1}, we do not show that this functional is bounded, and hence we cannot use measure theory.  This is different from the case of Bosonic quantum field theories, which correspond to measures on configuration spaces.\footnote{There is some resemblance to the case of purely Fermionic quantum field theories.}

We begin with a brief review of some results in Constructive Quantum Field Theory which provide both analogies and contrasts to the construction in this paper.  We refer the reader to \cite{GJ} for an overview of this field as well as for complete proofs.  We then sketch the methods required to deal with functional integrals of the type (\ref{ffi}), where reflection positivity is used as a substitute for measure theory.

\subsection{Some Results from Constructive Quantum Field Theory  }\labell{ccqft}
Choose a representation of $G$ giving a trace which produces a nondegenerate inner product on the Lie algebra $\fg.$\footnote{For simplicity, we may consider the case where $M$ is a rational homology sphere, $G$ is a compact classical Lie group, and the inner product is given by negative of the trace in the defining representation.}    The Chern-Simons invariant appearing in (\ref{ffi}) has the form
\begin{equation}\labell{CS}
{\rm{CS}}(A) = \frac{1}{4\pi}{\rm tr} \int_M A d A + \frac23 A^3
\end{equation}
It is therefore a sum of a quadratic polynomial and a higher degree polynomial in the connection form $A.$  Such expressions occur commonly in Quantum Field Theory.  One starting point in constructive quantum field theory is to consider a quadratic term as giving a Gaussian measure on some appropriate Banach space; the higher degree terms can be ``cut off'' to give functions on that space, and an appropriate limit of the integrals of these functions is the desired non-Gaussian measure.

Given a separable Hilbert space $\cH,$ there exists a Gaussian measure $d\mu$ on a Banach space $\cB \supset \cH$ characterized by the integrals of bounded functions $\exp(i h(\cdot))$, where $h \in \cB^\star \subset \cH^\star;$\footnote{Here we denote by $\cH^\star$ the Banach space dual of the Hilbert space $\cH.$  Of course, this is isomorphic to $\cH.$  In this paper, $\cH$ will be a Sobolev space of differential forms on some manifold $M,$ of Sobolev class $-s$, and $\cH^\star$ will be a similar Sobolev space of forms of class $s.$} namely,
\begin{equation}\labell{gaussian}
\int_{\cB} d\mu(\phi) e^{i h(\phi)} = e^{-\frac{1}{2} ||h||_{\cH^\star}^2};
\end{equation}
 
See e.g. \cite{gross}.  Equivalently, this measure may be characterized by its value on some unbounded, but integrable functions, of the form

$$p_{a_1,\dots,a_k}: \cB \to \R$$

\noindent where $a_1,\dots,a_k \in \cB^\star,$ given by

$$p_{a_1,\dots,a_k}(\phi) = a_1(\phi)a_2(\phi)\dots a_k(\phi);$$
\noindent we then have, for $k$ even,

\begin{equation}\labell{poly}
\int_{\cB} d\mu(\phi) p_{a_1,\dots,a_k}(\phi) = \frac{1}{2^{k/2} (k/2)!}\sum_{\sigma \in \Sigma_k} 
\langle a_{\sigma(1)}, a_{\sigma(2)} \rangle_{\cH^\star} \dots \langle a_{\sigma(k-1)}, a_{\sigma(k)} \rangle_{\cH^\star},
\end{equation}

\noindent where $\Sigma_k$ is the permutation group on $k$ letters. (If $k$ is odd, the integral is zero.)\\

This measure is an infinite dimensional analog of the finite dimensional Gaussian measure 

$$dm(x) = e^{-\frac{1}{2} ||x||^2} \frac{d^n x}{(2\pi)^{\frac{n}{2}}} $$

\noindent on a Euclidean space $\R^n,$ where $d^n x$ is Lebesgue measure; in this case the measure is defined on $\R^n$ and satisfies the analog of the property (\ref{gaussian}).  In infinite dimensions, neither the Lebesgue measure nor the function $e^{-\frac{1}{2} ||x||^2}$ has a reasonable analog, but the Gaussian measure has a good generalization, though only as a measure on a Banach space $\cB$ in which $\cH$ is a subspace of measure zero.  

The difficulties of quantum
field theory, involving the appearance of divergences in computations of expectations in such functional integrals, arise from the fact that the Banach space on which the measure lives is larger than the Hilbert space $\cH.$

More precisely, let $\cH$ be some appropriate Sobolev space of functions on a compact Riemannian manifold $M$ of dimension $d.$   In quantum mechanics and quantum field theory, the standard choice is the Sobolev space $\cH = H_{1}(M).$  Then the Banach space $\cB$ can be taken to be the Sobolev space $\cB = H_{1-\frac{d}{2}-\epsilon}(M)$ for any $\epsilon > 0.$  If $d = 1,$ this can be improved and the space $\cB$ can be taken to be a space of continuous functions of Lipschitz class ${\frac{1}{2}-\epsilon}.$\footnote{The precise choice of the Banach space $\cB$ does not play much of a role
in the construction.  One aspect of this fact is that if we consider two choices of Banach space given by $\cB = H_{1-\frac{d}{2}-\epsilon}(M)$ and $\cB^\prime = H_{1-\frac{d}{2}-\epsilon^\prime}(M),$ where 
$\epsilon > \epsilon^\prime,$
the open sets in $\cB^\prime$ arising from the inclusion $\cB^\prime \to \cB$ give a very different topology on $\cB^\prime$ than its native topology, but the same Borel sets, and therefore the same
measure theory.}
  If $d > 1,$ however, the distributions in $\cB$ are not functions, and polynomial functions of the type $\int_M \phi^k$ do not make sense as functions on $\cB.$

In the case $d = 1,$ where $M = S^1$ with the standard metric, this difficulty does not arise, and there is no difficulty
in defining the integral 

\begin{equation}\labell{1d}
Z_{QM}(\lambda) = \int_{\cB} d\mu(\phi) e^{-\lambda \int_M P(\phi(x)) dx}\end{equation}

\noindent where $P$ is any polynomial which is bounded below and $\lambda > 0.$  The Feynman-Kac formula then relates the integral $Z(\lambda)$ to Schrodinger operators on the real line.  We have 
 
\begin{equation}\labell{FK}
Z_{QM}(\lambda) =  C~ {\rm tr~} e^{ - (-\frac12\frac{d^2}{dx^2} + \frac12 x^2 + \lambda{P}(x))}
\end{equation}
 
\noindent where the trace is taken over $L_2(\R)$ and where $C$ is a constant independent of $\lambda.$\footnote{The quadratic term in the Schrodinger operator arises from the constant term added to the Laplacian in the definition of the Sobolev space.  This constant term has to be added to avoid problems with the constant mode in the Fourier expansion of $\phi,$  known in the Physics literature as an infrared divergence.  Such issues, though inessential, cause nontrivial difficulties we will have to surmount later in this paper.}  For now we note only that the Hilbert space on which the Schrodinger operator acts is not the Hilbert space $\cH$ used to define the path integral.  The difference between these two spaces will be important below.

In  the case of manifolds $M$ of dimension $d > 1,$ in order to overcome the problem in defining polynomial functions on the space $\cB,$ both formal quantum field theory and constructive quantum field theory
proceed by two steps:  Regularization and Renormalization.

For $\phi \in \cB$ and $\epsilon > 0,$ let $\phi_\epsilon \in C^\infty(M)$ denote the smooth function

$$\phi_\epsilon(x) = (e^{-\epsilon(\Delta + 1)}\phi)(x)$$

\noindent where $\Delta$ denotes the positive Laplacian on $M.$

We can use these smoothings of the distributions in $\cB$ to define polynomial functions on $\cB.$  To take the simplest example, the function $F_\epsilon: \cB \to \R$ given by the polynomial
\begin{equation}\labell{reg}
F_\epsilon(\phi) = \int_M \phi_\epsilon(x)^3 dx
\end{equation}

\noindent is a well defined function on $\cB.$  This is ``regularization''.

Now we wish to take $\epsilon \to 0.$  The family of functions $F_\epsilon$ are all square integrable with respect to the measure $d\mu,$ but they do not have a limit in $L_2(d \mu).$  However, let
\begin{equation}\labell{renorm}
 : F_\epsilon (\phi) : = F_\epsilon(\phi) - 3 \int_M C_\epsilon(x)  \phi_\epsilon(x) dx,
\end{equation}
\noindent where $C_\epsilon(x) =  (e^{-\epsilon (\Delta + 1)} \frac{1}{\Delta + 1} e^{-\epsilon (\Delta + 1)})(x,x) .$  This is "renormalization".

Suppose now that $d=2.$  A computation using (\ref{poly}) shows that

\begin{equation}\labell{cauchy}
||: F_\epsilon (\phi) : -: F_{\epsilon^\prime} (\phi) :||^2_{L_2(\cB,d\mu)} \to 0
\end{equation}
\noindent as $\epsilon, \epsilon^\prime \to 0.$\footnote{This amounts to convergence of the value of the $\Theta$ graph in second order perturbation theory.  This in turn follows from the fact that the Green's function for the operator $\Delta + 1$ on a compact two dimensional manifold has logarithmic singularities, and no worse, along the diagonal.} Hence the family $:F_\epsilon:$ converges in $L_2(\cB, d\mu)$ to a limit
function $:F:,$ and therefore the exponential

\begin{equation}\labell{interacting}
e^{i \lambda :F:}
\end{equation}
\noindent is a bounded function on $\cB,$ hence integrable, for any $\lambda \in \R.$  The resulting 
integral 
\begin{equation}\labell{interactingi}
\int_{\cB} d\mu(\phi) e^{ i \lambda :F(\phi):}
\end{equation}
\noindent is the mathematical expression corresponding to the formal functional integral
\begin{equation}\labell{interactingif}
\int d\phi~ e^{-\int_M (\frac12|d\phi(x)|^2 + \frac12 \phi(x)^2 - i \lambda \phi(x)^3) dx }.
\end{equation}
This formal expression, where the Lagrangian is the sum of a quadratic term and an imaginary cubic, may be viewed as a rough analog of the formal Chern-Simons functional integral (\ref{ffi}).

Returning for the moment to (\ref{interactingi}), we see that as long as we are interested in integrals of functions of the form (\ref{interacting})--in Physics language, with imaginary coupling constant--the existence of the limit function $:F:$, which is equivalent to the convergence of the expectation of the square of a  polynomial function (in Physics language, finiteness up to second order in perturbation theory), is sufficient to guarantee the existence of the functional integral {\em nonperturbatively}.\footnote{In particular, one formally infinite adjustment (\ref{renorm}) suffices to render all the terms in the formal perturbative series finite. It is perhaps possible to view the finiteness of the perturbation series for Chern-Simons gauge theory established by Axelrod-Singer \cite{as} also as arising from the nonperturbative construction.}

Note that formally, the function $:F:$ is a linear combination (\ref{renorm}) of polynomials with infinite coefficients.  This is an  example of the infinite constants associated to renormalization in quantum field theory.  However, mathematically, there is nothing mysterious about the limit in (\ref{cauchy}).
\footnote{If $d > 2,$ the limit in (\ref{cauchy}) does not exist, and more sophisticated methods are required.  See e.g. \cite{GJ3,Hai,Sokal,AizD}.}

\begin{Remark}  We have chosen to consider the case of a cubic polynomial with pure imaginary coefficient due to the resemblance to the Chern-Simons functional.  For purposes of the Feynman-Kac formula, it is necessary as in (\ref{FK}) to consider polynomials which are bounded below, such as quartics with a positive real coefficient.  Take $M = S^1 \times S^1$ in the flat metric.  If we take 

$$F_\epsilon(\phi) = \int_M \phi_\epsilon(x)^4 dx,$$

\noindent then the appropriate renormalization is 

$$:F_\epsilon(\phi): = \int_M \phi_\epsilon(x)^4 dx - 6 C_\epsilon \int_M \phi_\epsilon(x)^2 + 3 C_\epsilon^2.$$

\noindent (For $S^1 \times S^1$ in the flat metric, $C_\epsilon(x)$ is a constant we denote by $C_\epsilon.$)  Again, the limit  $:F: = \lim_{\epsilon \to 0} F_\epsilon$ exists by calculations using (\ref{poly}).
However, the term $- 6 C_\epsilon \int_M \phi_\epsilon^2 $ in the definition of $:F_\epsilon:$ is not positive, and consequently  the limit function $:F:$ is not bounded below.  Thus integrability of  the function $e^{-:F:}$  requires more delicate estimates; see \cite{GJ}, Section 8.6. These results are based on the fact that estimates on the measure of the set where $e^{-:F:}$ is large can be made in terms of estimates on integrals of polynomial functions, though more subtle than those of (\ref{cauchy}).  There are also entirely different problems addressed in Constructive Quantum Field Theory to obtain the infinite volume limit and prove the Wightman axioms.  None of these issues  arise in our construction of Chern-Simons gauge theory.\end{Remark}

\subsection{The Chern Simons functional integral}
 
We now outline the ideas required to construct a mathematical version of the Chern Simons functional integral.

Consider the Lagrangian (\ref{CS})

$$
{\rm{CS}}(A) = \frac{1}{4\pi}{\rm tr} \int_M A d A + \frac23 A^3,
$$

\noindent which appears in the formal functional integral (\ref{ffi})

$$
Z_k(M) = \int_\cA dA e^{i k {\rm CS}(A)}
$$
\noindent  multiplied by an imaginary factor $ik.$  We would like to proceed by analogy to the formal expression (\ref{interactingif})

\begin{equation*}
\int d\phi~ e^{-\int_M (\frac12|d\phi(x)|^2 + \frac12 \phi(x)^2 - i \lambda \phi(x)^3) dx }.
\end{equation*}

\noindent whose mathematically well-defined version is (\ref{interactingi})

\begin{equation*} 
\int_{\cB} d\mu(\phi) e^{ i \lambda :F(\phi):}.
\end{equation*}  

In (\ref{interactingif}), the quadratic term is real and the cubic is multiplied by an imaginary coupling.\footnote{This combination of a real, negative definite quadratic term and an imaginary cubic is necessary even in one dimension, to obtain convergent integrals over $\R$ giving rise to Airy functions.} The formal functional integral (\ref{ffi}) can also be brought into this form by a formal change of variable.  We therefore take as our starting point the formal functional integral

\begin{equation}\labell{fffi}
\int_{\cA} dA e^{  {\rm tr~} \int_M \frac{1}2 A dA + i \lambda A^3}
\end{equation}

\noindent where $\lambda \in \R.$   The fact that perturbative Chern-Simons gauge theory has been a useful approach is one piece of evidence in favor of studying this type of functional integral.
\footnote{
A similar phenomenon, where the mathematically well-defined functional integrals give the objects of physical interest by a form of analytic continuation, is well-understood in constructive quantum field theory.  Feynman's formal functional integrals
take the form

$$Z_{Formal}(\lambda) = \int d\phi e^{i\int_M \phi \square \phi + \lambda P(\phi)},$$

\noindent where $\square$ is the wave operator; whereas the non-Gaussian integrals of constructive quantum field theory correspond to formal expressions of the type
\begin{equation}\labell{pf} 
Z_{Euclidean}(\lambda) = \int d\phi e^{-\int_M \phi \Delta \phi + \lambda P(\phi)}.
\end{equation}

Thus the objects of study in the formal Feynman integrals might be expected to be obtainable from formally related but well-defined objects in the non-Gaussian measures corresponding to (\ref{pf}) by analytic continuation.  The issue of whether such an analytic continuation exists, and what properties it has, was extensively studied in constructive quantum
field theory; we refer to \cite{sw,GJ} for more information.
}

Consider then further the formal functional integrals (\ref{ffi}) and (\ref{fffi}).
To begin with, the quadratic term appearing in both is degenerate, with the
degeneracy corresponding to the gauge symmetry of the Chern-Simons function.
The method of dealing with this degeneracy is standard:  We choose a Riemannian metric on the three manifold
$M,$ and replace the space of all connections by the space of connections
on $M$ satisfying the 
gauge condition

\begin{equation}\labell{gc}
d^\star A = 0.
\end{equation}
(We continue to denote this space by $\cA.$)  In order to avoid further complications, we restrict our attention to compact,
connected manifolds with trivial real first homology group; i.e.

$$H^1(M,\R) = H^2(M,\R) =0$$
$$H^0(M,\R) = H^3(M,\R) = \R.$$

In order to have any chance of the formal functional integral yielding
topological invariants, the fact that the gauge choice (\ref{gc}) depends
explicitly on the metric must be addressed by including additional ghost
fields in the action.  This modification of the formal functional integral
(\ref{ffi}) was addressed in \cite{Witten,as}.  Following their methods, the formal functional integral we wish to address mathematically is

\begin{equation}\labell{start}
\int dA dc_0 dc_2 e^{ {\rm tr~} \int_M \frac{1}2
(A dA -  2c_0 dc_2) + i \lambda (A^3 - 6 A c_0 c_2)}
\end{equation}

\noindent where the connections $A$ satisfy the gauge condition (\ref{gc}) and
where the ghost field $c_0$ is a anticommuting scalar with values in $\fg$ lying in the orthogonal complement $\Omega^0(M,\fg)^\perp$ of the constants in $\Omega^0 (M,\fg) $\footnote{The orthocomplement of the constants can be taken since the constant ghost modes decouple in the free part of the action (\ref{start}).} 
and the ghost field $c_2$ is
an anticommuting field with values in $\Omega^2(M,\fg)$ and satisfying the gauge condition $d^* c_2 = 0.$   

Our aim then is to give precise meaning
to all the elements appearing in (\ref{start}).
For now we note that the formal Berezin integral

\begin{equation}\labell{berezin}
\frac1{Z_{ghost}}\int  dc_0 dc_2 e^{  -{\rm tr~} \int_M (c_0 d c_2)} (\cdot)
\end{equation}

\noindent (where $Z_{ghost}$ is a formally infinite normalization factor) is a mnemonic device denoting the pairing of
$\bigwedge^* (\Omega^0(M,\fg)^\perp)$ with $\bigwedge^* (\Omega^2(M,\fg)) $ induced by the inverse of the operator $\star d:\Omega^0(M,\fg)^\perp \to \Omega^2(M,\fg) $
and causes complications, but virtually no difficulties, in our construction.  The main task of this paper is to interpret formal functional
integrals of the type
\begin{equation}\labell{quad}
\frac1{Z_{gauge}}\int_{\cA} d A e^{\frac{1}2 tr \int_M A d A } (\cdot)
\end{equation}
\noindent (where $Z_{gauge}$ is a formally infinite normalization factor) as giving a linear functional with useful positivity properties on a subalgebra of the algebra of functions on $\cA.$  

Now the quadratic term in the exponential in (\ref{quad})
is not positive, and therefore does not correspond in a naive way to a 
Gaussian measure.  However, this quadratic term does possess a more subtle
form of positivity, known in constructive quantum field theory
as Reflection Positivity (or Osterwalder-Schrader Positivity) \cite{os}.

More explicitly, we will consider manifolds $M$ equipped with a diffeomorphism $R,$ which preserves, {\em but does not fix,}\footnote{Reflection positivity in
constructive quantum field theory generally considers reflections which {\em fix} a time-zero submanifold.  
In our case, the
nontrivial action of $R$ on $\Sigma$ is crucial to the construction of a positive definite inner product.} a compact, connected, oriented
submanifold $\Sigma$ lying in $M,$ and where $M - \Sigma$ consists of
two connected components $M_+$ and $M_- = R M_+.$
We require
the action of $R$ on $\Sigma$ to lift to an action on $H^1(\Sigma,\R)$ which
makes the bilinear form $Q: H^1(\Sigma,\R) \otimes H^1(\Sigma,\R) \to \R$
given by
\begin{equation}\labell{quadf}
Q(a,b) =-  \int_\Sigma R^*a \wedge b 
\end{equation}

\noindent {\em positive definite} on cohomology classes extending to $M_-.$\footnote{In fact, much less will suffice; we can consider any manifold formed by gluing two copies of a three manifold with boundary $M_+$ by a diffeomorphism of the boundary having 
the appropriate action on cohomology.  See Section \ref{mwr}.}
We then choose an $R-$invariant metric on $M$ and study expressions given by formal expectations of functions on $\cA$ given by 
a product of linear functionals on $\cA$ in the formal integral (\ref{quad}),
and given by

\begin{equation}\labell{quad1}
\frac{1}{Z_{gauge}}\int_{\cA} d A e^{\frac{1}2 tr \int_M A d A } \langle f, A\rangle \langle g,A\rangle =  \langle f, L g\rangle
\end{equation}

\noindent where $f, g \in \Omega^1(M,\fg)$ satisfy $d^*f = d^*g = 0,$  $L = (\star d) \Delta^{-1}$, $\Delta$ is the Laplacian on 
$\Omega^1(M,\fg)$, whose inverse exists due to the cohomological condition $H^1(M,\R) = 0,$ and the inner product $\langle \cdot, \cdot \rangle$ is the $L^2$ inner product on one forms with values in $\fg$ associated to the inner product on $\fg.$\footnote{In the technical part of the paper, once we have interpreted formal functional integrals in terms of appropriate spaces of connections, the inner products appearing on the left hand side of (\ref{quad1}) will be replaced by evaluation of  distributions on elements of their dual spaces.}  (Then $\frac12 {\rm tr}\int_M A d A = -\frac12 \langle A, \star dA \rangle$ and formal application of the rule (\ref{poly}) gives (\ref{quad1}).)  The main theorem is then

\begin{varTheorem} \labell{osp}

Suppose $f \in \Omega^1_c(M_+,\fg)$ is a $\fg$-valued one form compactly supported on $M_+,$ and satisfies $d^*f = 0.$
Then
\begin{equation}\labell{inp}
\langle R^\star f , Lf \rangle_{L_2} \geq 0\end{equation}
\end{varTheorem}

The positive semidefiniteness in (\ref{inp}) is the positive semidefiniteness of the expectations of certain quadratic polynomials in the
formal functional integral associated to (\ref{quad}), and implies a similar positivity property for the expectations of {\em any} polynomial
function arising from functions compactly supported on $M_+$ (Theorem \ref{rp}). Thus, the formal
expectations motivated by (\ref{inp}) give a reflection positive linear functional on 
a space of functions on a Banach space $\cbA$ of  connections on $M.$  In order to obtain a Hilbert space from this construction we must take a quotient by the those functions which are in the kernel of this reflection positive functional.  We will show in Theorem \ref{bhs} that the resulting
Hilbert space is given by 

\begin{equation}\labell{identi} H^B(\Sigma) = {\rm Sym}^\star(\Lambda)^- ,\end{equation}

\noindent where  $\Lambda \subset H^1(\Sigma,\fg)$ is the image of $H^1(M_-,\fg)$ in $H^1(\Sigma,\fg)$ and the symmetric algebra 
${\rm Sym}^\star(\Lambda)$ is completed in the norm arising from the bilinear form $Q$ in 
(\ref{quadf}) and the metric on $\fg.$  \footnote{In concrete terms, the identification in (\ref{identi}) is given by the map taking a one form $f \in \Omega^1_c(M_+,\fg)$ satisfying $d^*f = 0$ to the cohomology class $[Lf|_\Sigma] \in H^1(\Sigma,\fg).$  Since $f$ is compactly supported on $M_+,$ the one form $Lf$ is closed on $\Sigma,$ and extends to a closed form on $M_-;$ in other words, $[Lf|_\Sigma ] \in \Lambda.$}

Including the Fermions gives a full Hilbert space

$$H(\Sigma) = H^B(\Sigma) \otimes H^F(\Sigma)$$

\noindent  where $$H^F(\Sigma)=\bigwedge{}^{{}^\star} (H^0(\Sigma,\fg) \oplus H^2(\Sigma,\fg))$$  is a finite dimensional vector space.  Thus the Hilbert space is a symmetric algebra in (a subspace of) the even homology of $\Sigma,$ tensored with an alternating algebra on the odd cohomology of $\Sigma.$  A subtlety is that the positive definite metric on $H^F(\Sigma)$ is not the natural one from the point of view of topological quantum field theory; this is related to the fact that the Fermions are ghosts.  We will show how to deal with this issue.

So far we have dealt with the quadratic term in the Chern-Simons functional, which is the only term occurring in abelian Chern-Simons gauge theory, and our Reflection
positivity result falls naturally into what may be expected from such a theory.  It remains to deal with the cubic term appearing in 
(\ref{start}).  (We continue 
to focus on the space $\cbA$ of 
connections, and leave consideration of the ghosts for the technical part
of the paper.)  We wish to make sense of an expression of the form

\begin{equation}\labell{fpi}
e^{i \lambda tr \int_M A^3}.\end{equation}

Now unlike the case of cubic Bosonic quantum field theory, where such an expression can be interpreted as a function on the analog of the space of connections, our reflection positivity theorem only applies to polynomial functions\footnote{See Section \ref{Kto0} for a further discussion of this point.}, so we have to interpret (\ref{fpi}) in terms of  the Hilbert space $H^B(\Sigma).$

To do so, we write 

$${\rm tr} \int_M A^3 = \Xi_B^+  + R^* \Xi_B^+$$

\noindent where 

\begin{equation}\labell{xibp}
\Xi_B^+ = {\rm tr} \int_{M_+} A^3
\end{equation}

\noindent is a polynomial\footnote{The polynomial in (\ref{xibp}) looks like a cubic, but in fact is linear in each of the components of the gauge field, and is an exterior product of those components, so its singularity properties are much better than that of a cubic of the type appearing in (\ref{interacting}).  One aspect of this is the vanishing of the Wick ordering correction.  Since we cannot use this polynomial interaction, due to other reasons explained below, we do not attempt to study it further.} in the gauge field supported on $M_+.$  So morally, we may expect that the interaction term corresponds to a selfadjoint multiplication operator on $H^B(\Sigma).$

However, the polynomial $\Xi_B^+$ is not quite an element of the configuration space, since it is not a limit of polynomials in terms of the form $A(f)$ where $f$ is a one form compactly supported on $M_+$ {\em and satisfying the condition $d^*f = 0.$}  One way of seeing that this difficulty is essential is to note that the expectation in the formal path integral of the product of $\Xi_B^+$ with a cubic polynomial of the form $A(Rf) A(Rg) A(Rh),$ where $f, g, h \in \Omega_c^1(M_+,\fg)$ satisfy $d^*f = d^*g = d^*h = 0,$ depends on the closed one forms $Lf|_\Sigma, Lg|_\Sigma, Lh|_\Sigma,$ and not only on the cohomology classes (and Hilbert space images) $[Lf|_\Sigma], [Lg|_\Sigma], [Lh|_\Sigma];$ see (\ref{defint}).  So in order to define a Hilbert space element corresponding to $\Xi_B^+$ we have to make some choices of lifts of cohomology classes to elements of the configuration space, and the Hilbert space element $\xi_B^+$ corresponding to $\Xi_B^+$ will depend on these choices.

Having made these choices, we obtain a cubic element $\xi_B^+ \in H^B(\Sigma)$ of the Hilbert space, a corresponding densely defined symmetric operator $\xi_B^+\cdot +(\xi_B^+\cdot)^*$, a selfadjoint extension  $O_B = (\xi_B^+\cdot +(\xi_B^+\cdot)^*)^-,$
and a one parameter subgroup of bounded operators $e^{i\lambda O_B}$ on $H^B(\Sigma)\otimes \C.$  We may thus consider the "Bosonic partition function"

$$Z_B(\lambda) = \langle \Omega,  e^{i\lambda O_B} \Omega\rangle$$

\noindent where $\lambda \in \R$ and $\Omega$ is the image of the polynomial $1$ in $H^B(\Sigma).$

The inclusion of the Fermion ghosts causes complications of various sorts, since, as we noted, the positive definite metric for the Fermion Hilbert space  is not the correct one for topological field theory.  However, this ghost Hilbert space is finite dimensional, so this can be managed at the cost of some complications.  Again, the definition of the interaction term requires choices.  We can then ask whether the combined partition function is independent of the choices we made in the definition of $\xi_B^+$ and their equivalent for the Fermion interaction term.

It is also an open question whether this method of constructing a Chern-Simons interaction term is powerful enough to reproduce the manifold invariants of Witten and Reshetikhin-Turaev.  In standard quantum field theory, it is necessary to take a time ordered exponential of interacting terms, not an ordinary exponential of the type we have studied.  On the other hand, in a topological quantum field theory, where the Hamiltonian is zero and the effect of an interaction term depends only on topological data (see Appendix A for a manifestation of that), it is possible that this does not matter.  See Questions \ref{conj1}, \ref{conj2}, and \ref{conj3} for concrete statements.\footnote{A technical comment:  The invariants of Witten and Reshetikhin-Turaev are defined for framed manifolds.  Since we choose a metric on $M$ once and for all, and normalize the quadratic functional integral associated to this metric to have value $1$ on the polynomial $1,$ we are effectively normalizing the functional integral in a way that removes this framing dependence.}

The remainder of the paper is structured as follows.  In Section \ref{mwr}, we construct manifolds, which we require to be rational homology spheres, equipped with a reflection map adequate for an algebra of functions on the corresponding configuration spaces to carry a reflection positive functional.  We also study cases where these manifolds satisfy further conditions which allow a more geometric construction of these reflection maps.  In Section \ref{config}, we construct the configuration spaces for the gauge and ghost fields.  There is a great deal of freedom in constructing such configuration spaces, and we make a convenient choice.  We then proceed in Section \ref{ospb} with the main result of the paper (Theorem \ref{osp} above) which shows that the quadratic term in the Chern-Simons function gives a reflection positive functional on the one particle configuration space of the gauge fields, and more generally a reflection positive functional on the polynomial algebra of fields (Theorem \ref{rp}).  We also compute the Hilbert space which this reflection positive functional associates with a two manifold given morally by the fixed set of the reflection. In Section \ref{ospfs}, we study a similar construction for the Fermion, or ghost fields.  The key result is Theorem \ref{rp1pf}, followed by Proposition \ref{piisom}.  We then combine the results of Sections \ref{ospb} and \ref{ospfs} and summarize them in Theorem \ref{ospcombined}.  Finally in Section \ref{intsec} we construct operators on these Hilbert spaces corresponding to the interaction terms, and in Theorem \ref{pfdef} obtain the partition function as a weak limit of vacuum expectations values of exponentials of these operators. We end with a discussion of possible relations to the manifold invariants of Witten and Reshetikhin-Turaev.

\section{Manifolds with reflection}\labell{mwr}

Let $M_+$ be a smooth, compact, connected, oriented 3-manifold with boundary $\Sigma,$ a smooth, compact, connected, oriented 2-manifold of genus $g > 0.$
The orientation on $\Sigma$ gives the vector space $H^1(\Sigma,\R)$ a symplectic
structure $\omega$ given by the intersection form.  The image $\Lambda_+$ of $H^1(M_+,\R)$ in $H^1(\Sigma,\R)$ is a Lagrangian subvariety.

Let $S: \Sigma \to \Sigma$ be an orientation reversing diffeomorphism whose action $S^*$ on $H^1(\Sigma,\R)$ satisfies\footnote{Note that since the conditions (\ref{conds}) on $S$ involve only the action $S^*$ of $S$ on cohomology, and since the representation of the mapping class group in $Sp(2g,\Z)$ given by the action on cohomology is a surjection onto $Sp(2g,\Z)$ (see \cite{mcg}), any element of $Sp(2g,\Z)$ satisfying (\ref{conds}) arises from the action of some diffeomorphism.}
\begin{align}\labell{conds}\begin{split}
(S^*)^2 = & 1\\
\omega(S^*a,a) >  0 & {\rm ~for ~all~nonzero~} a  \in \Lambda_+.\end{split}
\end{align}
 
Then $\Lambda_\R = S^* \Lambda_+$ is also Lagrangian, and 
\begin{equation}\labell{intneg}
\omega(S^*a,a) <  0  {\rm ~for ~all~nonzero~} a  \in \Lambda_\R.\end{equation}

We can use $S$ to construct a compact, connected, oriented 3-manifold $M$ by gluing two copies of $M_+,$ which we denote by  $M_+, M_-,$ by the diffeomorphism $S$ of their common boundary.  From now on we impose the
assumption

$$ H^1(M,\R) = 0.$$

Our construction of the functional integral will work for any rational homology sphere arising from a gluing satisfying the conditions (\ref{conds}).  However, the construction will simplify considerably if the diffeomorphism $S$ satisfies some finiteness
conditions.  One such condition is
\\~\\
{\bf Condition A:}  $S$ is of finite order.
\\~\\
\indent It simplifies further if $S$ satisfies
\\~\\
{\bf Condition B:} $S^2 = {\rm id}.$
\\~\\
\indent If Condition B is satisfied, $M$ comes equipped with an orientation preserving diffeomorphism $R$, exchanging $M_+$ and $M_-,$ and whose action on $\Sigma$ is given by $S;$ this is the situation considered in the introduction.  To construct this orientation preserving diffeomorphism $R : M \to M,$ write
$$ M = (M_+ \amalg M_-)/\sim,$$

\noindent where $m_+ \in M_+, m_- \in M_-$ satisfy $m_+ \sim m_-$ if $m_+ \in \partial M_+ = \Sigma,$ $m_- \in \partial M_- = \Sigma,$ and $m_-= S m_+$ (as elements of $\Sigma$).  Then define $R_+: M_+ \to M_-$ by

$$ R_+(m) = j(m)$$

\noindent and $R_- : M_- \to M_+$ by 

$$R_-(m) = j^{-1}(m)$$

\noindent where 
\begin{equation}\labell{defj}
j: M_+ \to M_-
\end{equation}

\noindent is the identity diffeomorphism identifying the two copies $M_+,M_-$ of $M_+.$  Then $R_+$ agrees with $R_-$ on $\Sigma,$ and gives a diffeomorphism of $M$ preserving $\Sigma$ and whose action on $\Sigma$ is given by $S.$  Choosing an invariant metric this diffeomorphism gives an involution of $\Omega^*(M)$ preserving the de Rham operator and the Riemannian inner product, and exchanging the subspaces $\Omega^*_c(M_+)$ and $\Omega^*_c(M_-)$ of compactly supported differential forms.

We now choose a compact Lie group $G$ and a representation of $G$ whose associated trace gives a metric on $\fg.$ This metric, combined with the Riemannian metric, gives a Hodge star operator and a metric on the differential forms with coefficients in $\fg.$  The involution $R^*$ then extends to the differential forms with coefficients in $\fg$ and preserves the metric and Hodge star operator.
 
In the case of the weaker Condition A, we do not know how to construct such a diffeomorphism, but it is still possible to construct an
involution exchanging forms compactly supported on $M_+$ with those compactly supported on $M_-$ with the properties needed to define a satisfactory form of reflection positivity.  To do so, note that the identification $j : M_+ \to M_-$ induces an isomorphism $j^*: \Omega^*_c(M_-,\fg) \to \Omega^*_c(M_+,\fg)$ on the compactly supported forms.  To find a metric for which $j^*$ preserves also the Hodge $\star$ operator, 
choose an $S-$invariant metric $g$ on $\Sigma,$ and a metric on $M_+$ taking the form $g + (dt)^2$ on a collar neighborhood $U = \Sigma \times [0,1)$ of the boundary
$\Sigma.$  Gluing this metric to the same metric on $M_-$ gives a metric on $M$ which also has this collar form on a neighborhood of $\Sigma \subset M.$  Then the map $j^*$ satisfies 

$$j^* (\star \xi) = \star j^* (\xi)$$

\noindent for any $ \xi \in  \Omega^*_c(M_-,\fg).$  Since $j$ is orientation preserving, the map $j^*$ preserves the metric on compactly supported forms.  We continue to denote the resulting map by 

$$R^* : \Omega^*_c(M_\pm,\fg) \to \Omega^*_c(M_\mp,\fg),$$

\noindent  whether or not it arises from a diffeomorphism of $M.$ 

In the absence of Condition A, we can still obtain such a map as long as we restrict to forms compactly supported on the complement of a collar containing $\Sigma.$  To do so, we construct a metric on $M$ by again equipping $M_+$  with a metric taking the collar form $g + (dt)^2$ on a neighborhood of its boundary, and gluing $M_+$ to a copy of $M_-$ equipped with the same metric, using a collar $\Sigma \times [-\delta,\delta]$ equipped with a metric given by   $g + (dt)^2$ near $\Sigma \times \{-\delta\}$ and by  $S^*g + (dt)^2 $ near 
$\Sigma \times \{\delta\}.$  To perform the gluing, we use the identity on the boundary of $M_-$ and the diffeomorphism $S$ on the boundary of $M_+.$ We therefore obtain a metric on $M$ where the metrics on $M_-$ and $M_+$ agree on the complement of a $ \delta$ neighborhood $W_\delta$ of $\Sigma.$     Then the map $j$ still preserves the metric, and therefore the Hodge $\star$ operator, on the complement of $W_\delta.$  By restricting $j^*: \Omega^*_c(M_-,\fg) \to \Omega^*_c(M_+,\fg)$ to forms compactly supported away from this collar, we obtain a map

$$R^* : \Omega^*_c(M_\pm^\delta,\fg) \to \Omega^*_c(M_\mp^\delta,\fg),$$

\noindent where $M_\pm^\delta = M_\pm \cap (M - W_\delta).$  The involution $R^*$ preserves the Hodge $\star$ operator, and therefore the metric on forms.  It will then be necessary to take another
limit, as $\delta \to 0,$ of the partition function and other quantum field theoretic objects.  

In sum, we have chosen in all cases a Riemannian metric on $M,$ along with an invariant metric on $\fg$ arising from the trace in a representation.  These endow the differential forms with coefficients in $\fg$ with a Hodge star operator and a metric.  We also have a reflection $R^*$  which may or may not arise from the action of an involution on $M,$ but which exchanges the differential forms on $M$ compactly supported on $M_+^\delta = M_+ \cap (M-W_\delta)$ (where in the case of Condition A, we may take $\delta = 0$) with those compactly supported in its image in $M_-,$ and preserves the metric and star operator on such forms.\footnote{We consider forms compactly supported on $M_\pm$ as forms on $M$ by extension by zero.}

We will from now on consider mainly the case where Condition A is satisfied, and indicate the simplifications that occur
in the case of Condition B, and the additional constructions needed in the absence of either condition.  
\section{The configuration spaces $\cF^B$ and $\cF^F$}\labell{config}
In this section we construct the configuration spaces
of our quantum field model.  The configuration space $\cF^B$ for the gauge fields is
the completion of the symmetric algebra on the one-particle
configuration space $\cF_1^B.$  The space $\cF^B$ can also be viewed as a space of integrable functions 
on a space of connections $\cbA,$ with respect to a Gaussian measure we construct on $\cbA.$

Similarly, we construct a one-particle configuration space $\cF_1^F$ corresponding
to the ghost configuration space, and the completion $\cF^F$ of its alternating
algebra is the configuration space for the ghost fields.  

\subsection{The one-particle configuration spaces $\cF_1^B$ and $\cF_1^F$}\labell{31}

Consider the subspace $F_1^B= {\rm ker}(d^\star) \subset \Omega^1(M,\fg),$ and complete
it to a Hilbert space $\cF_1^B$ using the $H_s$ Sobolev norm on $\Omega^1(M,\fg),$ with $s$ to be chosen
later; we may take $s = 4.$\footnote{The precise choice of $s$ does not play much of a role in our construction, so long as it is chosen large enough.  It is convenient for de Rham theory and the Hodge theorem to have forms with two continuous derivatives, which is the case if $s > 7/2.$  We will need $s >\frac14$ for (\ref{hs}) and $s$ to be an integer for Remark \ref{multrem}.}
Since $H^1(M,\R) = 0,$ the Laplacian $\Delta = d d^\star + d^\star d$ on 
$\Omega^1(M,\fg)$ has an inverse $\Delta^{-1}$ which extends to a
bounded operator on $\cF_1^B.$  The operator 
$$L = \star d \Delta^{-1}$$
 \noindent is then also a bounded operator on $\cF_1^B.$  For any open set $U \subset M,$ there are subspaces $F_1^B(U)\subset F_1^B$ and  $\cF_1^B(U)\subset \cF_1^B$ consisting of forms compactly supported on $U.$

The odd variables, or Fermions, appearing in the action (\ref{start}) are a scalar $c_0 \in \Omega^0(M,\fg)$ and a two form
$c_2 \in \Omega^2(M,\fg).$ 
Consider first the space of smooth forms $\Omega^0(M,\fg).$  Inside this space
we may consider the orthocomplement of the constants 
$\fg \subset \Omega^0(M,\fg)$
in the $L^2$-inner product;
in other words the functions
 $f \in \Omega^0(M,\fg)$
 with $\int_M \star f = 0.$  We denote this subspace of $\Omega^0(M,\fg)$ by
$\Omega^{0}(M,\fg)^\perp.$  Similarly we consider the space ${\rm ker}(d^*_2) \subset \Omega^{2}(M,\fg).$
We then form the space $\cF^F_{1}$ by completing the space
$$F_1^F= \Omega^{0}(M,\fg)^\perp  \oplus {\rm ker}(d^*_2)$$

\noindent in the $H_s$\footnote{We again choose the same $s$ (i.e. $s=4$) for similar reasons to those in the Bose case.} Sobolev norm given by the metric on $M$ and the inner product on $\fg.$

On the space $F_1^F,$ the operator $\star d$ acts, interchanging the scalars and the two forms.  Since we have taken the 
orthocomplement of the constants in $\Omega^0$ and the kernel of $d_2^*$ in $\Omega^2,$ this
operator has an inverse (as in \cite{as}; see also \cite{adams}) $ L = \star d \Delta^{-1},$ which we continue to denote by $L,$  a bounded operator on $\cF^F_{1}.$

As in the Bose case, we may define a one particle configuration space corresponding to any open subset $U \subset M.$ There is a slight subtlety due to the fact that in the construction of $F_1^F$ we took the 
orthocomplement of the constants in the zero forms.  So let $F_1^F(U) = (a,b) \in \Omega^0(M,\fg)^\perp \oplus {\rm ker}({d_2^*}) $ with $da$ and $b$ compactly supported inside of $U,$ and let $\cF^F_1(U)$ be the completion of $F_1^F(U)$ in $\cF^F_1.$

\subsection{The Bose configuration space $\cF^B$}\labell{bfs}

We define the Bose configuration space by 

$$\cF^B = {\rm Sym}^\star(\cF_1^B)^-,$$

\noindent the completion of the symmetric algebra ${\rm Sym}^\star(\cF_1^B)$ on $\cF_1^B$ in the norm
induced on ${\rm Sym}^\star(\cF_1^B)$ by the Hilbert norm on $\cF_1^B.$  For future purposes we also define the dense subspace of polynomials
$\cF_P^B = {\rm Sym}^\star(\cF_1^B) \subset \cF^B;$   likewise the subspace $\cF_d^B = {\rm Sym}^{\leq d} (\cF_1^B) \subset \cF^B$ of polynomials of degree $\leq d.$

 \begin{Remark}\labell{girmk} The space $\cF^B$ has a description in terms of Gaussian integrals.

Let $\cbA $ be the space of $\fg$-valued one forms $A$ on $M$ of Sobolev class $-s -\frac32 - \epsilon$ for some $\epsilon > 0$ and satisfying the condition $d^*A=0.$  On the space $\cbA$ consider the the Gaussian measure $d \mu$ with covariance $\Delta^{s};$ again we may take $s=4.$\footnote{The formal path integral corresponds to the action $S(A) = \langle A ,\Delta^{-s} A\rangle_{L_2}.$} In the terms used in the introduction, this measure can be obtained from the Hilbert space $\cH^* = \cF_1^B$ by completing it to a Banach space $\cB = \cbA$; see \cite{gross}.

This measure is characterized by integrals of exponentials, or alternatively, polynomial
functions, on $\cbA:$  Given $f  \in \cbA^\star \subset \cH^\star ,$
 
\begin{equation}\labell{Gaussianexp}
\int_\cbA d\mu (A) e^{i f(A)} = e^{-\frac12\langle f, f\rangle_{\cH^\star}}
\end{equation}

\noindent and given $f_j \in \cbA^\star \subset \cH^\star  , j = 1,\dots, m,$ the polynomial function 
$F(A) = f_1(A)\dots f_m(A)$ is integrable, and for $m$ even
\begin{equation}\labell{gaussianpoly}
\int_\cbA d \mu (A) f_1(A)\dots f_m(A) =\frac{1}{2^{m/2}{(m/2)!} }\sum_{\sigma \in \Sigma_m} \langle f_{\sigma(1)}, f_{\sigma(2)} \rangle_{\cH^\star } \dots\langle f_{\sigma(m-1)},f_{\sigma(m)}\rangle_{\cH^\star }
\end{equation}
\noindent where $\Sigma_m$ is the permutation group on $m$ letters.  (For $m$ odd, the integral is zero.)\\
 
Then  $\cF_P^B$ is isometrically embedded in $L^2(\cbA, d\mu ).$\footnote{This embedding uses Wick ordering.  It is a bit simpler to build $\cF^B$ in the measure given by $d\mu$ on the complexification of ${\bf A},$ as the completion of the space of real polynomials considered as functions on the complexified fields, since this avoids Wick ordering troubles.}
\end{Remark}

In view of Remark \ref{girmk}, when convenient, we will sometimes consider elements of $\cF^B$ as functions on the space ${\bf A},$ interchangeably with their nature as elements of a completed symmetric algebra.  So for example, if $f_1,\dots,f_d \in F^B_1,$ the expression $f_1(\cdot) \dots f_d(\cdot)$ is the function on $\cbA$ associated to the element $f_1 \otimes \dots \otimes f_d \in \cF^B_P.$

The following dense subsets of $\cF^B$ will be convenient to keep in mind.
 
 {\em The ring of polynomials.}  Let $\cR = {\rm Sym}^\star(F_1^B).$

Equivalently, we may use a different set of polynomials, which is also dense:

{\em The L-Wick ordered polynomials.}  The $L$-Wick ordered polynomials $\cO$ are generated as a vector space by monomials of the form
\begin{equation}\labell{wickpoly}
:m_{f_1,\dots,f_m}: = : f_1(\cdot) \dots f_m(\cdot):,
\end{equation}

\noindent where $f_1,\dots,f_m \in F_1^B  $ as above.  This is a polynomial of degree $m$ with coefficients which can 
be extracted from the generating function
 
$$ : e^{i f(\cdot)}: = e^{\frac12 \langle f, L f \rangle_{L_2}} e^{i f(\cdot)}.$$

Since any polynomial can be written as a linear combination of $L-$Wick ordered polynomials, these are also a dense subset of $\cF^B.$\footnote{See further comments on $L-$ Wick ordered polynomials in the next Section.}

For future reference we note that the definition of $:m_{f_1,\dots,f_m}:$ can be extended also to the case where $f_i \in \cF_1^B.$

\subsection{The configuration space $\cF^B(U)$ associated to $U \subset M$}

Inside $\cF^B  $ we may consider the subspace $\cR(U) = {\rm Sym}^\star(F_1^B)(U),$ and likewise 
the subspace $\cO(U)$ generated by elements of the form 
(\ref{wickpoly}) where $f_1,\dots,f_m$ are smooth and have compact support
in $U.$  The completion of any of these sets in $\cF^B$
is a Hilbert space we denote by $\cF^B(U).$
This space may also be built up by completing the symmetric algebra of a
Hilbert space,  restricting to the case of one forms compactly supported
in $U:$  Letting $\cF_{1}^B(U) = \cF_1^B \cap \cF^B(U),$ we have 
$$\cF^B(U) = ({\rm Sym}^\star\cF_{1}^B(U) )^-$$

\noindent and we define

$$\cF^B_P(U) = \cF^B_P \cap \cF^B(U)$$ 
\noindent and

$$\cF^B_d(U) = \cF^B_d \cap \cF^B(U)$$

\begin{Remark}\labell{multrem} If we choose $s$ to be a nonnegative integer, the covariance in Remark \ref{girmk} is a differential operator, and has zero correlations between disjoint open sets.
Explicitly, if $U, V$ are disjoint open sets, and $P \in \cF^B_P(U), Q \in \cF^B_P(V),$ then $PQ \in \cF^B_P$ and

$$\int d\mu P^2 Q^2 = \int d\mu P^2 \int d\mu Q^2;$$
\noindent alternatively,

$$||PQ||_2 = ||P||_2 ||Q||_2.$$

In other words, 

\begin{Lemma}\labell{multop}
Multiplication by $P \in \cF^B_P(U)$ is a bounded operator taking $\cF^B_P(V)$ to $\cF^B_P.$
\end{Lemma} \end{Remark}
\subsection{The Fermionic configuration Space $\cF^F$}\labell{ffs}
Consider the alternating algebra $\cF_P^F= \bigwedge^\star \cF_{1}^F.$  The norm on $\cF_{1}^F$ makes $\cF_P^F$ into an inner product space, so it may
be completed to a Hilbert space we call $\cF^F .$  Note that multiplication by
an element of $ F_{1}^F$ is bounded on $F^F$ and
hence extends to a bounded operator on $\cF^F .$  (This is in contrast to the
case of $\cF^B,$ where multiplication by an element of $ F_1^B$ is not a 
bounded operator on ${\rm Sym}^\star(\cF_1^B),$ and hence does not extend
to an operator on ${\rm Sym}^\star(\cF_1^B)^{-} = \cF^B.$) 
 
\begin{Remark} Alternatively, the inner product on $\cF^F$ can be considered as a functional on the larger alternating algebra 
$\cU = \cF^F_L \otimes  \cF^F_R$  where $\cF^F_L, \cF^F_R$ are two copies of $\cF^F;$ the tensor product is the smallest alternating algebra containing both factors.  Then
any monomial $m \in \cU$ may be written as the product $ m = m_L m_R$ where $m_L$ is a monomial
in $\cF^F_L$ and $m_R$ is a monomial in $\cF^F_R.$  We then define the functional $< \cdot > : \cU \to \R$ by defining

\begin{equation}\labell{bzd} < m_L^r m_R> = \langle m_L, m_R \rangle \end{equation}

\noindent using the identification $\cF^F_L \simeq \cF^F_R = \cF^F$ and extending by linearity.  (Here the notation $m_L^r$ denotes the monomial $(-1)^{\frac{|m_L|(|m_L|-1)}{2}} m_L$, where $|m_L|$ is the number of terms in $m_L;$ morally $m_L^r$ is $m_L$ written in reverse order.)\footnote{A technical point:  There is no need to Wick order $m_L$ and $m_R$ on the left side of (\ref{bzd}) since in the Berezin integral (\ref{bzi}), any monomial of the form $m_L$ or $m_R$ is equal to its own Wick ordering.}

In the language of Berezin integrals, the functional $<\cdot>$ is given by the Berezin integral 

\begin{equation}\labell{bzi}< p > = \int d c_{0,L} d{c}_{0,R} d c_{2,L} dc_{2,R} e^{-{\rm tr} \int_M \star {c}_{0,L}\Delta^{-s} c_{0,R} + \star {c}_{2,L} \Delta^{-s} c_{2,R} } p,\end{equation}

\noindent where $p$ is any polynomial in the odd variables $c_{0,L}, c_{0,R}, c_{2,L}, c_{2,R},$ which can be thought of
as an element of $\cU.$  These Berezin integrals are analogs of the Gaussian integrals of Remark \ref{girmk}. \end{Remark}

As in the case of $\cF^B,$ we may consider the subalgebra
$\cF^F(U) \subset \cF^F$ corresponding to an open set $U\subset M.$   Let $\cF_P^F(U) = \bigwedge^* \cF_1^F(U) \subset \cF^F,$  and let $\cF^F(U)$ be the completion of $\cF_P^F(U)$ in $\cF^F.$  Likewise we define $\cF^F_d(U)$ as the inhomogeneous polynomials of degree $\leq d.$
 
 We now write\footnote{The tensor products below are tensor product of Hilbert spaces.}

$$ \cF = \cF^B \otimes \cF^F$$

\noindent and, for any subset $U \subset M,$

$$ \cF(U) = \cF^B(U)  \otimes \cF^F(U)$$

\noindent as well as

$$ \cF_d(U) = \bigcup_{k + l = d} \cF_k^B(U)  \otimes \cF_l^F(U)$$

\noindent and 

$$ \cF_P(U) =  \cF_P^B(U)  \otimes \cF_P^F(U).$$

A special case is that where $U = M_+^\delta;$ we write $\cF^B_+ = \cF^B(M_+^\delta),$ $\cF^F_+ = \cF^F(M_+^\delta),$ and $\cF_+ = \cF (M_+^\delta)$ (where in the case of condition A or condition B being satisfied, we replace $M_+^\delta$ by $M_+.$). 
We will write  $\cF^B_{P,+} = \cF^B_P \cap \cF^B_+,$ $\cF^F_{P,+} = \cF^F_P \cap \cF^F_+,$ $\cF_{P,+} =  \cF_P\cap \cF_+,$ and $\cF^B_{d,+} = \cF^B_d \cap \cF^B_+,$ $\cF^F_{d,+} = \cF^F_d \cap \cF^F_+,$ $\cF_{d,+} =  \cF_d\cap \cF_+.$  Likewise we write $\cR_+$ and $\cO_+.$

\section{Reflection Positivity and the Bosonic Hilbert Space}\labell{ospb}

The formal functional integral 

$$\int_{\cA} e^{\frac{i}{2}  {\rm tr} \int_M A d A} \cdot$$

\noindent does not correspond in any known way to a measure on a Banach space of connections.
However, it turns out that the expectations derived from application of the formal rule of the type (\ref{poly}) to polynomial functions in the formal functional integral (\ref{quad})

\begin{equation}\labell{ffifree}\frac1{Z_{gauge}}\int_{\cA} e^{\frac{1}2 {\rm tr} \int_M A d A} (\cdot)\end{equation}
\noindent give rise to a reflection positive functional on the space of polynomials, in the following sense:  Suppose that $P,\tilde{P}$ are polynomials  compactly supported in $M_+.$\footnote{In this informal discussion, we consider the case where Condition A is satisfied.}  Then application of the rules of the type (\ref{poly}) to the formal expression
$$q (P,\tilde{P}) = \frac1{Z_{gauge}}\int_{\cA} e^{\frac{1}2 {\rm tr} \int_M A d A} (R^*P) \tilde{P}$$
\noindent gives a positive semidefinite symmetric bilinear form on the space of polynomials.
   We may construct the Hilbert space $H^B(\Sigma)$ by taking the quotient of the space of polynomials by the null elements, and completing in the norm induced by the bilinear form $q.$
Along the way we prove that 

$$H^B(\Sigma) = ({\rm Sym}^\star \Lambda)^{-}$$

\noindent where $\Lambda \subset H^1(\Sigma,\fg)$ is the image of $H^1(M_-,\fg)$ in $H^1(\Sigma,\fg),$ and the closure is taken in the norm on the symmetric algebra ${\rm Sym}^\star \Lambda$ arising from the norm
on $\Lambda$ given by the symmetric positive definite bilinear form

$$Q(a,b)   =  {\rm tr} \int_\Sigma S^*a \wedge b, $$

\noindent for $a, b \in \Lambda.$

\subsection{Reflection positivity for the one particle configuration space $\cF_{1,+}^B$}

The following theorem is the main idea in the construction of our reflection positive functional.  Recall that $\Delta$ is the Laplacian on differential forms, whose 
restriction to one-forms is invertible, and that $L = \star d \Delta^{-1}$ is a compact operator.
We now prove Theorem \ref{osp} from the introduction.

\begin{Theorem}\labell{rp1p}  Let $f \in \Omega^1(M,\fg)$ be a smooth $\fg$-valued one
form on $M$ compactly supported in $M_+$ (in the absence of Condition A, $M_+^\delta$) and satisfying $d^* f = 0.$ Then

\begin{equation}\labell{rp1pe}
\langle R^*f , Lf \rangle_{L_2} \geq 0\end{equation}
\end{Theorem}

{\em Proof:}
We first consider the situation where $M$ comes equipped with a reflection $R$ (Condition B).
We then have 
\begin{align}\begin{split}\labell{osppf}
\langle R^*f , Lf \rangle_{L_2} =
- {\rm tr}\int_M R^*f \wedge  \star L f =\\
- {\rm tr}\int_{M_-}R^*f \wedge \star Lf = 
- {\rm tr}\int_{M_-} dLR^*f \wedge Lf =\\
 -{\rm tr}\int_{M_-} LR^*f \wedge \star f - {\rm tr}\int_{-[\Sigma]} R^*Lf \wedge Lf .
\end{split}\end{align} 
\noindent where we recall the orientation of $\Sigma$ is that given by the orientation of $M_+;$ this gives rise to the negative sign in orientation of $\Sigma$ in the final equality in (\ref{osppf}).  Since $f$ is compactly supported on $M_+,$

$${\rm tr}\int_{M_-} LR^*f \wedge \star  f = 0.$$

\noindent

But $\star dLf = f,$ and therefore $dLf$ is compactly supported on $M_+.$
Hence $Lf|_\Sigma$ is a closed $\fg$-valued one form on $\Sigma,$ which extends to a closed one form $Lf|_{M_-}$ on $M_-;$ in other words $[Lf|_\Sigma] \in \Lambda=\Lambda_\R \otimes \fg .$ Then
 
$${\rm tr}\int_{\Sigma} R^*Lf \wedge Lf  = {\rm tr} \int_\Sigma R^*[Lf|_\Sigma] \wedge [Lf|_\Sigma],$$

\noindent where $[Lf|_\Sigma] $ is the cohomology class of $Lf|_\Sigma$ in $H^1(\Sigma,\fg).$
But the bilinear form $Q:H^1(\Sigma,\fg) \otimes H^1(\Sigma,\fg) \to \R$ given
by 

\begin{equation}\labell{ourq}Q(a,b) = {\rm tr} \int_\Sigma S^*a \wedge b\end{equation}

\noindent is symmetric and positive definite on $\Lambda= \Lambda_\R \otimes \fg$ by (\ref{intneg}).\footnote{Note that the trace differs from the positive inner product on the Lie algebra by a sign, so that the fact that the bilinear form in (\ref{intneg}) is negative definite makes the form $Q$ in (\ref{ourq}) positive definite.} Since  $R|_\Sigma = S,$ 

$${\rm tr} \int_\Sigma R^*[Lf|_\Sigma] \wedge [Lf|_\Sigma] \geq 0.$$

In the more general case (Condition A), a slightly more elaborate argument is needed, but the essentials are the same.  We begin with a Lemma.\footnote{Note that in the case where $g = R^*f,$ Lemma \ref{L to coho} agrees with the result of Theorem \ref{rp1p}.}
 
\begin{Lemma}\labell{L to coho}
Let $f \in \Omega^1_c(M_+,\fg)$ and $g\in \Omega^1_c(M_-,\fg).$  Then 

$$ {\rm tr}\int_M  f \wedge \star L g ={\rm tr} \int_\Sigma Lf \wedge Lg.$$
\end{Lemma}

\begin{pf}

We have 
\begin{align}\begin{split} {\rm tr}\int_M f \wedge \star L g = {\rm tr}\int_{M_+}  f \wedge \star L g = {\rm tr}\int_{M_+} \star d Lf \wedge \star Lg = \\
 {\rm tr}\int_{M_+} L f\wedge g + {\rm tr}\int_\Sigma Lf \wedge Lg.\end{split}\end{align}
But $g$ is compactly supported in $M_-,$ so $ {\rm tr}\int_{M_+} L f\wedge g =0.$\end{pf}

Note that (in the notation of the above Lemma) $d L f|_\Sigma = d Lg|_\Sigma = 0,$ so that $Lf$ and $Lg$ are closed forms on $\Sigma.$  We have shown
that ${\rm tr}\int_M f \wedge \star L g = {\rm tr} ([Lf|_\Sigma] \cdot [Lg|_\Sigma])$ where $\cdot$ denotes
the intersection form.

In fact, Lemma \ref{L to coho} applies to any splitting of $M$ into two components by any submanifold $\tilde{\Sigma} \subset M.$  

\begin{Proposition}\labell{splitM}
  
Let $\tilde{\Sigma} \subset M$ be a compact connected oriented submanifold of $M$ whose complement has two components $M_1,M_2.$  Orient $\tilde{\Sigma}$ as the boundary of $M_1.$  Then if $f_1 \in F^B_1(M_1),$ $f_2 \in F^B_1(M_2),$ we have

$$ {\rm tr}  \int_M f_1 \wedge \star L f_2 ={\rm tr} \int_{\tilde{\Sigma}} [Lf_1|_{\tilde{\Sigma}}] \wedge [Lf_2|_{\tilde{\Sigma}}].$$

Furthermore, if $\Sigma_1, \Sigma_2$ are two isotopic compact submanifolds of $M,$ and $f$ is zero in the region between $\Sigma_1$ and $\Sigma_2,$ then $[Lf|_{\Sigma_1}] = [Lf|_{\Sigma_2}]$ where the two cohomology groups are identified by the isotopy. \footnote{ This is an indication that this theory has zero Hamiltonian.}

\end{Proposition}
We now continue studying the case where $M$ is split by our original $\Sigma.$
\begin{Lemma}\labell{raction}
Let $f \in \Omega^1_c({M_+},\fg).$  Then 

$$ [LR^* f|_\Sigma] = S^* [Lf|_\Sigma].$$
\end{Lemma}
 
\begin{pf} Denote by $\phi_\pm \in \Omega^1(M_\pm,\fg)$ the differential forms obtained by applying the maps $j$ and $j^{-1}$ exchanging $M_+$ and $M_-$ to the forms $Lf|_{M_\mp}.$   Note that these are not compactly supported forms.  Then we have

\begin{equation}\labell{excg} d L R^*f|_{M_\pm} =  \star R^*f|_{M_\pm}  = d \phi_\pm.\end{equation}

We wish to show that for any cohomology class $\xi \in H^1(\Sigma,\R)$

$$ \int_\Sigma ([(LR^*f)|_\Sigma] - S^*[Lf|_\Sigma]) \wedge \xi = 0.$$

Let $\delta > 0$ be sufficiently small so that $f$ is supported in $M^\delta_+.$ Let $\Sigma_\pm = \partial M^\delta_{\pm}.$  It will suffice to prove that there exists a basis $\{\xi_i\}$ for $H^1(\Sigma,\R)$  so that for each $i,$

$$ \int_{\Sigma_+}  ([(LR^*f)|_{\Sigma_+}] - [\phi_+|_{\Sigma_+}]) \wedge \xi_i  = 0 ~~{\rm ~~or~~} \int_{\Sigma_- }([(LR^*f)|_{\Sigma_-}] - [\phi_-|_{\Sigma_-}])\wedge \xi_i  = 0.$$
  
Let $i_\pm: \Sigma\to M_\pm$ denote the inclusion, and let $\Lambda_\pm = i_\pm^*( H^1(M_\pm,\R) )\subset H^1(\Sigma, \R).$   Since $H^1(M,\R) = 0,$ $\Lambda_+ \cap \Lambda_- = \{0\}$ and 
$\Lambda_+ \oplus \Lambda_- = H^1(M,\R).$  We claim that for any $\xi \in \Lambda_\pm,$ there exists $g \in \Omega^1_c(M_\mp,\R)$ with $d^* g = 0$ so that $[Lg|_\Sigma] = \xi.$\footnote{We define the operator $L$ on $\R$-valued one forms satisfying the gauge condition in the same way as for Lie algebra valued forms.}  To see this, we take a form $\tilde{\xi} \in \Omega^1(M,\R)$ which is closed on $M_\pm$ and on a neighborhood of $\Sigma,$ and such that $[\tilde{\xi}|_\Sigma] = \xi;$ such a form exists by the definition of $\Lambda_\pm.$  By the Hodge theorem, there exists $\psi \in C^\infty(M)$ with $d^* ( \tilde{\xi} + d\psi ) = 0.$  Let $g = \star d( \tilde{\xi} + d\psi).$  Then $Lg = \tilde{\xi} + d\psi,$ so that $[Lg|_\Sigma] = \xi.$  And $g = \star d \tilde{\xi},$ so that $g$ is compactly supported on $M_\mp.$

Take a basis $\{\xi_i^\pm\}$ for $\Lambda_\pm$ and choose forms $g_i^\pm \in \Omega^1_c(M_\mp,\R)$ with $d^*g_i^\pm =0$ and $[Lg_i^\pm|_\Sigma] = \xi_i^\pm$ for each $i.$  Take $\delta$ sufficiently small so that all the $g_i^\pm$ (as well as $f$) are supported in $M^\delta_\mp.$  Consider first the case of the $g_i^-.$  For each $i,$ we wish to show  that

$$\int_{\Sigma_-} (LR^*f - \phi_-) \wedge Lg_i^-=0.$$

To see this, compute

$$ \int_{\Sigma_-} (LR^*f - \phi_-) \wedge Lg_i^- = \int_{M^\delta_-} (d LR^*f - d\phi_-) \wedge 
Lg_i^- -  (LR^*f - \phi_-) \wedge \star g_i^-.$$

But $g_i^-|_{M_-} = 0,$ and $(d LR^*f)|_{M_-} - d\phi_- = 0$ by (\ref{excg}).  

A similar argument, substituting $\Sigma_+$ for $\Sigma_-,$ applies for the $g_i^+.$\footnote{In this case both of the relevant integrals vanish.}

In the absence of Condition A, the same result applies to forms compactly supported on $M^\delta_+.$

\end{pf}

{\em End of Proof of Theorem \ref{rp1p}:}  The proof of Theorem \ref{rp1p} in the case of Condition A now follows from Lemma \ref{L to coho} and Lemma \ref{raction}, along with condition (\ref{intneg}).  In the general case, where no finiteness condition is imposed on $S,$ the same theorem holds restricting to forms compactly supported on the complement $M_+^\delta$ of a collar neighborhood of $\Sigma.$  \hspace{.1in}$\square$

We now form a Hilbert space using the bilinear form given by Theorem \ref{rp1p}.
Recall first that the space $F_1^B(M_+)= {\rm ker~} d^* \subset \Omega^1_c(M_+, \fg)$ of 
smooth one-forms on $M$ compactly supported on $M_+$  is dense
in $\cF_{1,+}^B.$   (In the absence of Condition A, replace $M_+$ by $M_+^\delta.$)  Let $q: \cF_{1,+}^B \otimes \cF_{1,+}^B \to \R$ be the  
bilinear form given by 

$$q(f,g) = \langle R^*f , Lg \rangle_{L_2}.$$  Then $q$ is a positive
semidefinite bilinear form on $\cF_{1,+}^B.$

 Let $\cN^B_1 \subset \cF_{1,+}^B$
denote the space of null elements of $q.$  Then since $q(f,f) \leq ||f||_{\cF^B_{1,+}},$ $\cN^B_1$ is a closed subspace
of $\cF_{1,+}^B,$ and the quotient $\cF_{1,+}^B/\cN^B_1$ can be completed in
the norm given by $q$  to give a Hilbert space $H^B_1(\Sigma).$
The computation in the proof of Theorem \ref{rp1p} shows there is a close
relation between elements of $\cF_{1,+}^B$ and elements of $\Lambda \subset H^1(\Sigma,\fg).$

As in the proof of Theorem \ref{rp1p} we can construct a map $\pi_\Sigma^B :F_1^B(M_+)\to H^1(\Sigma,\fg)$ given by

$$\pi_\Sigma^B(f) = [Lf|_\Sigma].$$

\begin{Proposition}\labell{h1f1}
The map $\pi_\Sigma^B$ carries $ F_1^B(M_+)$
surjectively onto the image $\Lambda \subset H^1(\Sigma,\fg)$ of $H^1(M_-,\fg)$ in $H^1(\Sigma,\fg),$
and carries the bilinear form  $q$ on  $F_1^B(M_+)  $ to the positive definite bilinear form $Q$ on $\Lambda \subset H^1(\Sigma,\fg).$
(In the absence of Condition A, the same result holds with $M_+$ replaced with $M_+^\delta.$)
\end{Proposition}

\begin{pf} Let $f, g$ be two smooth $\fg$-valued one forms on $M$ compactly
supported on $M_+.$  By Lemma \ref{L to coho} and Lemma \ref{raction}, we have $q(f,g) = Q(R^*[Lf|_\Sigma], [Lg|_\Sigma]).$  Thus
the map $\pi_\Sigma^B:{\rm ker}(d_1^*) \subset \Omega^1(M_+,\fg) \to H^1(\Sigma,\fg)$ given by $\pi_\Sigma^B(f) = [Lf|_\Sigma]$ carries the 
bilinear form $q$ (restricted to smooth forms) to the bilinear form $Q.$

To prove that the image of $\pi_\Sigma^B$ lies in $\Lambda,$ note that if $f$ is a $\fg$-valued one form compactly supported on $M_+,$ the one form $Lf$ is closed on $M_-.$  Thus the class $[Lf|_\Sigma]$ extends to a class $[Lf|_{M_-}].$

To prove surjectivity, we proceed as in the proof of Lemma \ref{raction}.  Suppose $\xi$ is a closed $\fg$-valued one form on $\Sigma$ extending to $M_-$ as a closed form.  Then $\xi$ extends to a $\fg$-valued one form $\tilde{\xi}$ on all of $M$ whose restriction to $M_-$ is closed;  we may also take $\tilde{\xi}$ closed in a neighborhood of $\Sigma.$ By the Hodge Theorem and the vanishing of $H^1(M,\R),$ there exists a $\fg$-valued function $\phi$ such that
$d^* (\tilde{\xi} + d\phi) = 0.$  Then  $L \star d (\tilde{\xi} + d \phi) = \tilde{\xi} + d \phi,$  so that 

$$[L \left( \star d (\tilde{\xi} + d \phi) \right)|_\Sigma] = [\tilde{\xi}|_\Sigma + d(\phi|_\Sigma)] = [\xi].$$

\noindent Since $\star d (\tilde{\xi} + d\phi) = \star d \tilde{\xi} $ is supported in $M_+,$ $[\xi]$ is in the image of $\pi_\Sigma.$

\end{pf}

\begin{corollary} \labell{1phs} The map $\pi_\Sigma^B $  extends to a map $\pi_\Sigma^B: \cF_{1,+}^B \to H^1(\Sigma,\fg)$ and descends to an isomorphism  $\pi_\Sigma^B : \cF_{1,+}^B/\cN^B_1 \to H^B_1(\Sigma) \simeq \Lambda$ carrying the positive definite form induced on $\cF_{1,+}^B/\cN^B_1$ by $q$ to the positive definite form $Q$ on $\Lambda.$
\end{corollary}
 
\begin{Remark}
In view of Proposition \ref{splitM}, this construction can be performed for any splitting of $M$ by a submanifold $\tilde{\Sigma}$ isotopic to $\Sigma$ into two components $M_1,M_2.$ We obtain will a map $\pi_{\tilde{\Sigma}}^B: \cF^B_1(M_i)\to H^1(\tilde{\Sigma},\fg),$ 
$i = 1,2.$ Orient $\tilde{\Sigma}$ as the boundary of $M_1.$
The map  $\pi_{\tilde{\Sigma}}^B: \cF^B_1(M_i)\to H^1(\tilde{\Sigma},\fg),$ 
$i = 1,2;$ is an surjection onto the Lagrangian subspaces $\Lambda$ and $S^* \Lambda,$ respectively.  We use the notation $H^B_1(\tilde{\Sigma}) := \Lambda$ and continue to write $\pi_{\tilde{\Sigma}}^B: \cF^B_1(M_1) \to H^B_1(\tilde{\Sigma}),$ and similarly for the map $\pi_{\tilde{\Sigma}}^B: \cF^B_2(M_2) \to S^* \Lambda \simeq H^B_1(\tilde{\Sigma}).$

\begin{Proposition}\labell{anytwosplit} If $f_1 \in \cF^B_1(M_1), f_2 \in \cF^B_1(M_2),$ then 

$$\int_M \star f_1 \wedge L f_2 = \langle \pi_{\tilde{\Sigma}}^B (f_1), S^* \pi_{\tilde{\Sigma}}^B (f_2)\rangle_{H^B_1(\tilde{\Sigma})}.$$

Furthermore, if $\Sigma_1, \Sigma_2$ are two disjoint isotopic submanifolds of $M,$ splitting $M$ into three components, $M_1, M_2$ and $M_3= \Sigma \times [0,1], $ with $\partial M_3 = \Sigma_1 \cup \Sigma_2$, then if $f$
is compactly supported in $M_1$ (or $M_2$), then $$\pi_{\Sigma_1}(f) = \pi_{\Sigma_2}(f).$$
 \end{Proposition}
 \end{Remark}

\subsection{Reflection positivity for $\cF_+^B$}

We have now proved that the bilinear form $q$ gives a reflection positive
functional on $\cF_{1,+}^B.$  The form $q(a,b)$
gives the expectation of the product of the linear functions on $\cbA$ associated to the one forms
$a$ and $b$ in the formal functional integral (\ref{ffifree})

In this section we show that reflection positivity extends to $\cR_+, $ and hence to $\cF^B_{P,+}.$
To make a sensible definition corresponding to the formal functional integral
(\ref{ffifree}), we define a functional $\Phi : \cR \to \R.$ 

\begin{Definition}\labell{phi}  Let $\Phi : \cR \to \R$ be the linear functional
which is zero on all monomials of the form $f_1\otimes \cdots \otimes  f_n \in \cR$  (where $f_1,\dots,f_n \in F_1^B$) where $n$ is odd, and whose value on a monomial $f_1\otimes \cdots \otimes  f_n \in \cR$  (where $f_1,\dots,f_n \in F_1^B$) where $n$ is even is given by
\begin{equation}\labell{defphi}
\Phi(f_1\otimes \cdots \otimes  f_n)  = \frac{1}{2^{n/2}(n/2)!}\sum_{\sigma \in \Sigma_n} \langle f_{\sigma(1)}, Lf_{\sigma(2)}\rangle_{L_2} \dots \langle f_{\sigma(n-1)}, Lf_{\sigma(n)}\rangle_{L_2}.
\end{equation}

 \noindent Here the inner product $\langle \cdot, \cdot \rangle_{L_2}$ is the $L_2$ inner product of $\fg$-valued one forms on $M.$
\end{Definition}

\noindent  This functional corresponds to the formal functional integral 

$$ \frac1{Z_{gauge}}\int_\cA e^{ \frac{1}2 {\rm tr} \int_M A d A} f_1(A)  \dots f_n(A) d A$$  

\noindent where $Z_{gauge}$ is a formally infinite normalization constant.

\begin{Remark} Note also that we may scale the metric by a positive factor $\alpha$ (corresponding to a scaling of the quadratic term in the free Lagrangian in (\ref{quad})) to obtain a functional $\Phi_\alpha$ obtained by scaling the right hand side of (\ref{defphi}) by a factor of $\alpha^{-\frac{n}2}.$\end{Remark}

Alternatively, we may use the fact that, given our choice of $s=4,$ the operator $\Delta^{-s/2} L \Delta^{-s/2}$
is symmetric and Hilbert Schmidt,\footnote{For this property $s > \frac14$ would suffice.} as an operator on the $L_2$ completion $\overline{\cF_1^B}$ of $\cF_1^B$,  and hence gives a symmetric element $L_s \in \overline{\cF_1^B}\otimes \overline{\cF_1^B},$  to write 

\begin{equation} \labell{hs} \Phi(f_1\otimes  \dots\otimes  f_n ) = \frac{1}{2^{n/2}(n/2)!}\ \langle L_s \otimes \dots \otimes L_s, \Delta^{s/2} f_1\otimes \dots \otimes \Delta^{s/2} f_n\rangle_{{\rm Sym}^*\overline{\cF_1^B}}\end{equation}

\noindent Thus, given an inhomogeneous polynomial $P = \sum_l P_l$ of degree $d= 2k,$ where $P_l$ is homogeneous of degree $l,$ we may write 

$$\Phi(P) = \langle \sum_{j=0}^k \frac{1}{2^{j}j!} L_s^{\otimes j},   \sum_l (\Delta^{s/2})^{\otimes l} P_l   \rangle_{{\rm Sym}^*\overline{\cF_1^B}}$$

\noindent which shows that $\Phi$ extends to a bounded 
linear functional on $\cF^B_{d}  $ for any $d:$  that is, for $P \in \cF^B_{d},$

\begin{equation}\labell{bosebdd}|\Phi(P)| \leq C_d ||P||_{\cF^B_{d}}\end{equation}
 
\noindent for some constant $C_d$ which depends on $d.$  And $\Phi$ gives a (possibly-unbounded) linear functional on $\cF^B_P$ (see Section \ref{Kto0}).

We next define a bilinear form on $\cR_+$ which, since it extends the form $q$ on linear polynomials, we continue to denote by $q: \cR_+ \otimes \cR_+ \to \R,$ by

\begin{equation}\labell{qonpoly} 
q(P,Q) = \Phi((R^*P) Q).
\end{equation}

The bilinear form $q$ then also extends to $\cF^B_{P,+}$ and is bounded on any $\cF^B_{ d,+}.$

We prove the following

\begin{Theorem}\labell{rp}
Let $P \in \cR_+.$ 

Then $$q(P, P) \geq 0.$$
\end{Theorem}

Since $\cR_+\cap \cF^B_{d,+}$ is dense in $\cF^B_{d,+}$ for any $d,$\footnote{Recall that $\cF^B_{d,+}$ are the inhomogeneous polynomials of degree $d.$} this same result holds for $\cF^B_{d,+}.$  
We denote the nullspace of the bilinear form $q$ on $\cF^B_{d,+}$ by $\cN^B_d.$   The space $\cN^B_d$ is a closed subspace of $\cF^B_{d,+}$ by boundedness of $\Phi$ on $\cF^B_{d,+}.$  The quotient 
$\cF^B_{d,+}/\cN^B_d$ comes equipped with a bilinear form we also denote by $q$ which is positive definite,
and hence can be completed to a Hilbert space $H^B_{\leq d}(\Sigma).$  The resulting map $\pi_\Sigma^B : \cF^B_{d,+} \to H^B_{\leq d}(\Sigma)$ is bounded.  Writing $H^B_{d}(\Sigma) = H^B_{\leq d}(\Sigma)/H^B_{\leq d-1}(\Sigma),$ taking direct sums and writing
$H^B_P(\Sigma) = \oplus_d H^B_d(\Sigma),$ we obtain a map
$\pi_\Sigma^B:\cF^B_{P,+} \to H^B_P(\Sigma),$ which is bounded in any given degree.  The space $H^B_P(\Sigma)$ can in turn be completed to a Hilbert space $H^B(\Sigma).$
We will show below (see Proposition \ref{ident}) that $H^B_P(\Sigma) = {\rm Sym}^* \Lambda \subset {\rm Sym}^* H^1(\Sigma,\fg).$
 
The proof of Theorem \ref{rp} requires the following Lemma, giving reflection invariance of the Wick ordering for forms compactly supported on on $M_+$ (or $M_+^\delta,$ in the case where Condition A is absent).

\begin{Lemma}\labell{rinv}

Let $f \in \Omega^1(M, \fg)$ satisfy $d^*f = 0$ and be compactly supported on $M_+.$ Then

$${\rm tr} \int_{M} f \wedge \star L f = {\rm tr} \int_M R^*f \wedge\star L R^*f.$$

(In the absence of Condition A, replace $M_+$ by $M_+^\delta.$)
\end{Lemma}

\begin{pf} In the case of Condition B, this follows from the invariance of the metric under the involution $R.$\footnote{Note that for this invariance of the Wick ordering under reflection, it is crucial that the reflection $R$ is orientation preserving, and hence preserves the propagator $L.$  This is in contrast to the form of reflection positivity in the one particle space in \cite{w1}, where the reflection is orientation reversing and reverses the sign of the propagator.}

In the case of Condition A, we may not have a globally defined involution $R$; we have only an identification $j$  of $M_+$ with $M_-.$  But write $\psi = Lf,$ and define $\phi \in \Omega^1(M_+,\fg)$ by  $\phi =R^* ((LR^*f)|_{M_-});$
\footnote{This is a slight abuse of notation.  What we mean by $\phi =R^* ((LR^*f)|_{M_-})$ is the form $ \phi \in \Omega^1(M_+,\fg)$ (not of compact support) obtained from the form $LR^*f|_{M_-}\in \Omega^1(M_-,\fg)$ via the identification $j$ of $M_+$ with $M_-$ in (\ref{defj}).  We will only ever consider $\phi$ restricted to the support of $d\phi = \star f$ (which is also the support of $d\psi$), so no problems arise.} note that $\phi$ is not a form with compact support.  Then

\begin{equation}\labell{ds}
d \psi|_{M_+} =  \star f|_{M_+} = d \phi.\end{equation}

Thus

$$ {\rm tr}\int_M \star f \wedge L f = {\rm tr} \int_{M_+} \psi \wedge d \psi$$

and

$$ {\rm tr} \int_M \star R^*f\wedge L R^*f = {\rm tr} \int_{M_+} \phi\wedge d \phi.$$

Then

\begin{equation}\labell{peq}{\rm tr} \int_{M_+} \psi \wedge d \psi - \phi\wedge d \phi = {\rm tr} \int_{M_+} (\psi-\phi)\wedge d \psi + {\rm tr} \int_{M_+} \phi\wedge d (\psi - \phi).\end{equation}

The second term on the right hand side of (\ref{peq}) vanishes by (\ref{ds}).  The first term is given by
 
$$ {\rm tr} \int_{M_+} (\psi-\phi)\wedge d \psi = {\rm tr} \int_{M_+} d(\psi-\phi)\wedge \psi - {\rm tr}  \int_\Sigma [(LR^*f - R^*Lf)|_\Sigma]\wedge S^*[Lf|_\Sigma].$$

But $d(\psi-\phi) = 0$ by equation (\ref{ds}), and $[(LR^*f - R^*Lf)|_\Sigma] = 0$ by Lemma \ref{raction}.

In the absence of Condition A, replace $M_+$ by $M_+^\delta$ and the argument goes through as before.

\end{pf}

We now return to the Proof of Theorem \ref{rp}.
\begin{pf} Since any polynomial in $\cR_+$ can be written as a linear combination of 
$L$-Wick ordered polynomials in $\cO_+,$ it suffices to prove reflection
positivity for $L$-Wick ordered polynomials.  By Lemma \ref{rinv}, for any $L$-Wick ordered monomial $:m_{f_1,\dots,f_n}:$ where $f_1, \dots, f_n \in F^B_1(M_+),$

$$R^*:m_{f_1,\dots,f_n}: = :m_{R^*f_1,\dots,R^*f_n}:.$$

Then Definition \ref{phi} shows that for any $g_1,\dots,g_m \in F^B_1(M_+),$ the expression

$$\Phi( R^* :m_{f_1,\dots,f_n}: :m_{g_1,\dots,g_m}:)$$

\noindent vanishes if $m \neq n$ and is given by\footnote{Some comments about Wick ordering:  This is most convenient to explain for the case of Hilbert Schmidt operators.  Suppose $L$ is a symmetric Hilbert Schmidt operator on a Hilbert space $H.$  Then there exists $v \in {\rm Sym}^2 H \subset H\otimes H$ with

\begin{equation}\label{eq0}\langle v, x \otimes y\rangle_{{\rm Sym}^2 H} = \langle x, L y\rangle\end{equation}

for all $x,y \in H.$

Suppose $P$ is a polynomial in ${\rm Sym^*}H$ and define

$$\Phi(P) = \langle e^{i_v} P,1\rangle_{{\rm Sym^*}H}$$

where $i_v$ is interior product in ${\rm Sym^*}H;$ that is, 

$$i_v(x \otimes y) = \langle v, x \otimes y\rangle_{{\rm Sym}^2 H}.$$

Note that if $L > 0,$ the functional $\Phi$ has an alternative definition in terms of Gaussian integrals, as Gaussian measure with covariance $L,$ where the Hilbert space $\cH^*$ of (\ref{gaussian}) is the completion of $H$ in the norm given by $\langle \cdot,L \cdot \rangle.$ The definition we have given for $\Phi$ does not require positivity, although where positivity is not present, $\Phi$ does not correspond to a measure, and is defined as it stands only for polynomials.

Write, for $P \in {\rm Sym}^* H,$

$$:P: = e^{-i_v} P.$$

Then if $P,Q \in {\rm Sym}^* H,$

\begin{equation}\label{eq1}\Phi(:P::Q:) = \langle P, (\sum_{k=0}^\infty \frac1{k!}{\rm Sym}^k L )Q \rangle_{{\rm Sym}^*H}. \end{equation}

If $L$ is not Hilbert Schmidt, similar formulas hold as long as both sides of equation (\ref{eq1}) are well defined.  The Wick ordering can be defined by using equation (\ref{eq0}) to replace the interior product $i_v$ with an appropriate sequence of inner products.  Equation (\ref{eq1}) can be proved by cutting off $L$ to obtain a sequence of finite rank operators, and then taking the limit on both sides.  In our case an alternative argument uses transfer of regularity and the Hilbert Schmidt operator $L_s$ of equation (\ref{hs}).}

\begin{equation}\labell{phisym}
\Phi( R^* :m_{f_1,\dots,f_n}: :m_{g_1,\dots,g_m}: )=
\langle R^*f_1 \otimes \dots \otimes R^*f_n,   L g_1 \otimes \dots \otimes   L g_n\rangle_{L_2} \end{equation}

\noindent if $ m = n;$ here the inner product is the inner product on the symmetric algebra coming from $L_2$ inner product on $\fg$ valued forms. Therefore,  the 
inner product $q$ on polynomials in $\cO_+$ is the same as the inner product on ${\rm Sym}^*\cF^B_{1,+}$ induced by the inner
product given by $q$ on $\cF^B_{1,+}.$  It 
is therefore positive semidefinite.

In the absence of Condition A, the same result holds, for forms compactly supported on the complement $M_+^\delta$ of a collar neighborhood of $\Sigma.$

\end{pf}

In fact we have proved

\begin{Proposition}\labell{ident} The Hilbert space $H^B_d(\Sigma)$ is given by

$$H^B_d(\Sigma) \simeq {\rm Sym}^d H^B_1(\Sigma).$$

\end{Proposition}
 
\begin{pf}  The quadratic form associated to $q$ restricted to $\cF_{1,+}^B$ defines a bounded, self adjoint, nonnegative operator on $\cF_{1,+}^B.$  We can therefore find an orthonormal basis for $\cF^B_{1,+}$ consisting of a finite number of smooth forms $k_1,\dots,k_g$ forming a basis for the orthocomplement $H^B_1(\Sigma)$ of the kernel $\cN_1^B,$ and orthogonal also for the bilinear form $q,$ alongside an orthonormal basis of smooth forms $\{l_i\}_{i=1}^\infty$ for $\cN^B_1.$    This gives us an orthonormal basis for each ${\rm Sym}^d(\cF^B_{1,+})$ consisting of the elements $\{l_j \otimes p\},$ where $p=r_1\otimes \dots\otimes r_{d-1}$ is any monomial of degree $d-1$ in the $k_i$'s and $l_i$'s (i.e. $r_i \in \{k_1,\dots,k_g, l_1, \dots \}$ for all $i$), alongside the monomials $k_{i_1}\otimes\dots \otimes k_{i_d}.$  This basis is not orthogonal for $q.$  However, this basis for each ${\rm Sym}^{d^\prime}(\cF^B_{1,+})$ yields a basis for ${\rm Sym}^{\leq d}(\cF^B_{1,+})$ given by the elements $:m_{l_i,r_1,\dots,r_{d^\prime}}:,$ where $d^\prime \leq d,$  and where again $r_i \in \{k_1,\dots,k_g, l_1, \dots \}$ for all $i,$  alongside the monomials $:m_{k_{i_1}\dots k_{i_{d^\prime}}}:,$ where $d^\prime \leq d.$  This basis is orthogonal for the bilinear form $q_d=q|_{{\rm Sym}^{\leq d}(\cF^B_{1,+})}.$  By (\ref{phisym}), the bilinear form $q_d$ has nullspace given by the span of the $:m_{l_i,r_1,\dots,r_{d^\prime}}:,$ and is nondegenerate, and equal to the bilinear form induced by $q_1$ on ${\rm Sym}^{\leq d} H^B_1(\Sigma),$ on the span of the $:m_{k_{i_1}\dots k_{i_{d^\prime}}}:.$  The result follows.
\end{pf}

In fact, since we have seen in Corollary \ref{1phs} that there is an isometry $\cF^B_{1,+}/\cN^B_1 \simeq \Lambda,$ we have shown that there is an isometry

$${H}^B(\Sigma) = ({\rm Sym}^* \Lambda )^-,$$

\noindent identifying $H^B(\Sigma)$ with the completion of the symmetric algebra ${\rm Sym}^* \Lambda \subset  {\rm Sym}^* H^1(\Sigma, \fg)$\footnote{It is also possible to consider ${\rm Sym}^* \Lambda$ as an abelian subalgebra of the Weyl algebra on the vector space $H^1(\Sigma, \fg)$ equipped with the symplectic form $\tilde{Q}$ given by the intersection form on $\Sigma$ and defined in Remark \ref{qtilrmk} below; the subalgebra is abelian because $\Lambda$ is Lagrangian in this symplectic form.} in the norm induced on it by the norm on $\Lambda$ arising from inner product coming from the orientation of $\Sigma,$ the inner product on $\fg$, and the diffeomorphism $S:$ that is the inner product

$$Q(a,b) =  {\rm tr} \int_\Sigma S^*a \wedge b$$

\noindent for $a,b \in \Lambda.$  In terms of this identification, the map $\pi_\Sigma^B$ is given on the dense subset of $\cF^B_{d,+}$ (for each $d$) given by Wick ordered polynomials by writing\footnote{Note that the map $\pi_\Sigma^B$ is a filtered map, but not a graded map, of polynomials.}

\begin{equation}\labell{psexp}\pi_\Sigma^B ( : m_{f_1,\dots, f_{d^\prime}}:) = [Lf_1|_\Sigma] \otimes \dots \otimes [Lf_{d^\prime}|_\Sigma]\end{equation}

\noindent for each $d^\prime \leq d$ and extending by linearity.  The isomorphism induced by the map $\pi_\Sigma^B$  then carries the norm on of $H^B_d(\Sigma)$ isometrically to the norm on $ {\rm Sym}^d \Lambda .$  In summary we have proved

\begin{Theorem}\labell{bhs} The map $\pi_\Sigma^B$ given by (\ref{psexp}) descends to an isomorphism 

$${H}^B(\Sigma) \simeq ({\rm Sym}^* \Lambda )^-$$

\noindent where  $({\rm Sym}^* \Lambda )^-$ is the completion of ${\rm Sym}^* \Lambda$ in the norm induced on ${\rm Sym}^* \Lambda$ by the positive definite bilinear form ${\rm Sym}^*(Q).$

\end{Theorem}

The identification of ${H}^B(\Sigma) $ with the completion of a symmetric algebra endows ${H}^B(\Sigma)$ with a densely defined, self adjoint, nonnegative operator $N_B$ on $H^B(\Sigma)$ given by the polynomial degree, with domain $\cD = ({\rm Sym}^*\Lambda) \subset H^B(\Sigma).$ 

\begin{Remark}\labell{bscaling}In these terms the scaling of the quadratic term in the free Lagrangian by a factor of $\alpha$ (equivalent to replacing $\Phi$ by $\Phi_\alpha$) is equivalent to replacing the metric $\langle\cdot ,\cdot\rangle$ on the symmetric algebra by the metric $\langle\cdot, e^{-\log \alpha N_B} \cdot\rangle.$\end{Remark}

\begin{Remark}\labell{qtilrmk} In view of Proposition \ref{splitM}, we can find a formula similar to (\ref{phisym})  for any splitting 
of $M$ by a compact connected submanifold $\tilde{\Sigma}$ into two components $M_1,M_2.$  Define $\tilde{Q} : H^1(\tilde{\Sigma},\fg) \otimes H^1(\tilde{\Sigma},\fg) \to \R$ by

$$\tilde{Q} (a,b) =  {\rm tr} \int_{\tilde{\Sigma}} a \cup b.$$

Note that if $\tilde{\Sigma} = \Sigma,$ $\tilde{Q} (a,b) = Q(S^*a,b).$\footnote{The form $\tilde{Q}$ is closely related to the Atiyah-Bott symplectic form on the space of connections on $\Sigma.$}

\begin{Proposition}\labell{splitMpoly}
  
Let $\tilde{\Sigma} \subset M$ be a compact connected submanifold of $M$ whose complement has two components $M_1,M_2.$  Orient $\tilde{\Sigma}$ as the boundary of $M_1.$  Then if $:m_{f_1,\dots,f_n}:  \in \cO(M_1) $ and $:m_{g_1,\dots,g_n}:  \in \cO(M_2),$we have

$$ \Phi(:m_{f_1,\dots,f_n}: :m_{g_1,\dots,g_n}:)  = 
{\rm Sym}^*\tilde{Q}( [Lf_1|_{\tilde{\Sigma}}]\otimes\dots\otimes  [Lf_n|_{\tilde{\Sigma}}], [Lg_1|_{\tilde{\Sigma}}]\otimes \dots \otimes [Lg_n|_{\tilde{\Sigma}}])$$

\noindent 

\end{Proposition}

We therefore define $\hat{\pi}_{\tilde{\Sigma}}^B: \cF^B_P(M_1) \oplus \cF^B_P(M_2) \to {\rm Sym}^* (H^1(\tilde{\Sigma},\fg))$
by taking 

$$\hat{\pi}_{\tilde{\Sigma}}^B(:m_{f_1,\dots,f_n}: ) = [Lf_1|_{\tilde{\Sigma}}]\otimes \dots \otimes[Lf_n|_{\tilde{\Sigma}}]$$

and 

$$\hat{\pi}_{\tilde{\Sigma}}^B(:m_{g_1,\dots,g_n}: ) = [Lg_1|_{\tilde{\Sigma}}]\otimes \dots \otimes[Lg_n|_{\tilde{\Sigma}}]$$

\noindent where $:m_{f_1,\dots,f_n}:  \in \cO(M_1) $ and $:m_{g_1,\dots,g_n}:  \in \cO(M_2),$ and extending by linearity to  $\cF^B_P(M_1) \oplus \cF^B_P(M_2).$  We then have the following

\begin{Proposition}\labell{SplitM}  Let $\tilde{\Sigma}$ be any submanifold of $M$  isotopic to $\Sigma$ as oriented manifolds, and splitting $M$ into two components $M_1, M_2.$  Assume the orientation of $\tilde{\Sigma}$ agrees with that it inherits as the boundary of $M_1.$ Then if $P  \in \cF^B_P(M_1), Q \in \cF^B_P(M_2),$ 

$$\Phi(PQ) = \langle  (\hat{\pi}_{\tilde{\Sigma}}^B  (P)),  S^*\hat{\pi}_{\tilde{\Sigma}}^B (Q) \rangle_{H^B({\Sigma})}$$

\noindent where elements of ${\rm Sym}^*\Lambda \subset {\rm Sym}^*H^1(\tilde{\Sigma},\fg)$ are identified with elements
of $H^B(\Sigma)$ using the isotopy. \end{Proposition} 
 \end{Remark}

\subsection{Is there a relation between $\cF_+^B$ and ${ H}^B(\Sigma)?$}\labell{Kto0}
We consider the following question.

{\bf Question:}  Is there a bound 
\begin{equation}\labell{bd}
 |\Phi(a) | \leq C ||a||_{\cF^B} 
\end{equation}
\noindent for all $a \in \cR_+$?
 
In quantum field theories with polynomial interactions in two dimensions, such a bound is immediate, but in our case, where the functional $\Phi$ is
not defined using the measure $\mu $ of Remark \ref{girmk}, the existence of such a bound, especially since the operator $L$ is not positive, is not so
easy to see:  A positive linear functional on a $\C^*$ algebra is bounded, but I do not know of a comparable statement for a reflection positive functional.\footnote{The functional $\Phi$ can be extended to an algebra of bounded functions on ${\bf A}$ by taking imaginary exponentials of linear functionals.   For such functions arising from exponentiation of linear functions compactly supported on $M_+,$ $||\overline{(Rf)} f||_\infty = ||f||_\infty^2,$ which
resembles a $\C^*$ condition.  Another possible point of reference is Tomita-Takesaki theory.} The existence of such a bound 
(and the corresponding bound for Fermions) would
simplify considerably the proof of the existence of the interacting theory: The elements $e^{i \lambda \Xi_{B}^+}$
in $\cF_+\otimes \C$ would immediately descend\footnote{There is an additional point of concern about interpreting $\Xi_{B}^+$ as an element of $\cF_+;$  see Section \ref{intsec}.} to elements of $H(\Sigma) \otimes \C,$  as would their counterparts for $M_-;$ in fact, this would
hold for any $M_+, M_-$ having common boundary $\Sigma,$ without any need for a reflection.  This would lead to the existence of
partition functions in manifolds without reflection, and to strong bounds on those partition functions.   It would be worthwhile looking into
this as a guide to what kind of bounds, whether (\ref{bd}) or some weaker version, may be plausible.
Thus there is some interplay between bounds of the form (\ref{bd}) and bounds on partition functions in topological quantum field theory.  One possibility is that growth properties of manifold invariants which would preclude such a bound may contradict the existence of certain Heegaard splittings, so there is a potential for topological consequences.

One way to approach this question would be to replace the measure
$d\mu$ on ${\bf A}$ in Remark \ref{girmk}  with a measure $d\mu_K$ with covariance $L + K I,$ where $K \in \R$ satisfies $K > ||L||.$  The covariance $L + K I$ is positive, and the addition of $L$ is inconsequential for the singularities of the fields, so
it yields a measure on a Banach space of connections ${\bf A}$ corresponding to the case $s=0$ (instead of $s = 4$) in the construction of Remark \ref{girmk}; this space carries the White Noise measure on the space of connections. But the covariance $L + K I$ is also reflection positive, since 
\begin{equation}\labell{Kosp}
\langle R^*f, (L + KI) f \rangle = \langle R^*f, Lf \rangle
\end{equation}

\noindent for $f$ compactly supported in $M_+$; and (\ref{Kosp}) shows that the Hilbert space obtained from $d\mu_K$ by taking fields compactly supported on $M_+$ and then taking the quotient by the null vectors for the reflection positive functional is our same ${H}^B(\Sigma).$  Thus, replacing the functional $\Phi$ by the expectation in the measure $d\mu_K,$ the analog of bound (\ref{bd}) is satisfied, and we obtain a map from the configuration space of fields compactly supported on $M_+$ to ${H}^B(\Sigma).$ 

However, expectations in this new functional do not correspond in any way which I understand to Chern-Simons gauge theory; and I do not know how to take the
limit $K \to 0.$
 
 \section{The Fermion Hilbert space}\labell{ospfs}

In this section we study the Fermion Hilbert space.  Since the Fermions are ghosts, we may expect that positivity might not be straightforward, and there are in fact subtleties.  The Fermion Hilbert space  $H^F(\Sigma)$ is finite dimensional, so those subtleties can be overcome.

We will see that as a vector space,  $H^F(\Sigma) = \bigwedge^* (\fg \oplus \fg),$  and therefore $H^F(\Sigma)$ has the structure of an alternating algebra, which we can think of as the alternating algebra on the even cohomology $H^0(\Sigma,\fg) \oplus H^2(\Sigma,\fg),$ just as the Bosonic Hilbert space was a symmetric algebra on (part of) the odd cohomology.  This alternating algebra has a positive definite metric arising from the positive definite metric on $\fg \oplus \fg.$  This positive definite form does arise from a reasonable form of reflection positivity on the linear part of the configuration space.  But the required "twist" behaves badly under Wick ordering, so we do not get a reasonable form of reflection positivity for polynomials.    On the other hand, the nondegenerate split linear bilinear form on $\fg \oplus \fg$ does arise from reasonable formulas on the configuration space.\footnote{Note that it is conceivable that a subalgebra of the polynomials in the Fermi fields satisfies some form of reflection positivity even in the split linear metric.  One example of such a phenomenon is given in \cite{jp}.  I was not able to find such a subalgebra rich enough to contain the interaction term, and it is conceivable that due to the role of the ghosts in the interaction, no such subalgebra containing the interaction term can exist.  It would be interesting to investigate the type of Spin-Statistics Theorems (see e.g. \cite{sw}) that apply in this situation.}

Since  $H^F(\Sigma)$ is finite dimensional, the two inner products are related by a linear transformation; this gives a bounded operator on  $H^B(\Sigma)\otimes H^F(\Sigma),$ allowing us to work with Hilbert spaces when needed in treating the interaction term.

\subsection{The one particle configuration space $\cF^F_{1 }(U)$}

Recall that the one particle configuration space $\cF^F_{1}(U)$ associated to 
an open set $U\subset M$ was given by the completion of the space

$$F_1^F(U) = \Omega^0(U,\fg)^\perp \oplus \left( {\rm ker}(d^*_2) \cap \Omega^2_c(U,\fg) \right)$$

\noindent where the space $\Omega^0(U,\fg)^\perp$ is  the space of functions on $M$
which are constant outside a compact subset of $U,$ and whose integral over $M$ is zero.  Our first theorem is the analog of Theorem \ref{rp1p} for Fermions.

Suppose $\tilde{\Sigma}$ is a compact connected oriented submanifold of $M$, splitting $M$ into two connected components $U, V.$ 
Let $f \in \Omega^0(U,\fg)^\perp$ and $g \in {\rm ker} (d^*_2) \cap \Omega^2_c(V,\fg).$  Then

\begin{align}\begin{split}\labell{csts}
f|_{M - U } = c_f\\
Lg|_{M - V} = c_g\end{split}
\end{align}

\noindent where $c_f$, $c_g$ are constants.
\begin{Theorem}\labell{rp1pf}  Let $\tilde{\Sigma}$ be a compact connected oriented two-dimensional submanifold of $M,$ and suppose $M - \tilde{\Sigma}$ consists of two components $U,V.$  Let $f \in \Omega^0(U,\fg)^\perp$ and $g \in {\rm ker} (d^*_2) \cap \Omega^2_c(V,\fg),$ with $c_f, c_g$ as above.

Then 
\begin{equation}\labell{thm4eq}
{\rm tr}\int_M \star f  \wedge L g = - {\rm vol}(M) {\rm tr} c_f c_g.
\end{equation}

\end{Theorem}

\begin{pf} We have
\begin{align}\begin{split}
 {\rm tr} \int_M \star f \wedge L g =  {\rm tr}\int_{U} \star f c_g + {\rm tr} \int_{V} c_f \star Lg =\\
  {\rm tr}c_g (-\int_{V} \star f) + {\rm tr} c_f(- \int_{U} \star Lg),\end{split}\end{align}

\noindent using the fact that $\int_M \star f = \int_M \star Lg = 0.$ 
  
  But
  
 $${\rm tr}c_g (-\int_{V} \star f) + {\rm tr} c_f(- \int_{U} \star Lg) = -{\rm vol}(M) {\rm tr}c_f c_g$$
 
 \noindent as needed.
  \end{pf}

We now define $\pi^F_{\tilde{\Sigma}}: F^F_1(U) \to \fg \oplus \fg$ by

$$\pi^F_{\tilde{\Sigma}}(f,g) = (c_f, c_g)$$

\noindent where as in (\ref{csts})
\begin{align}\begin{split}
f|_{M - U } = c_f\\
Lg|_{M - U} = c_g,\end{split}
\end{align}

\noindent and likewise $\pi^F_{\tilde{\Sigma}}: F^F_1(V) \to \fg \oplus \fg$ by

$$\pi^F_{\tilde{\Sigma}}(f,g) = (c_f, c_g)$$

\noindent where again

\begin{align}\begin{split}
f|_{M - V } = c_f\\
Lg|_{M - V} = c_g.\end{split}
\end{align}

Then the map $\pi_{\tilde{\Sigma}}^F$ extends to $\cF^F_1(U),\cF^F_1(V).$  \footnote{ Where there is no danger of confusion, we denote $(f,0) \in F^F_1(M)$ simply by $f,$ likewise  $(0,g) \in F^F_1(M)$ simply by $g.$}  And if $df$ and $g$ are compactly supported
in the connected components of $U,V$ of $M- \tilde{\Sigma}$ as above,

$$ {\rm tr} \int_M \star f \wedge L g = \langle \pi_{\tilde{\Sigma}}^F(f), \pi_{\tilde{\Sigma}}^F(g) \rangle_{\fg \oplus \fg}$$

\noindent where the inner product on $\fg \oplus \fg$ is the split linear inner product\footnote{Recall that the inner product on $\fg$ is given by  $\langle a,b\rangle_\fg = -{\rm tr} (ab) $ for $a,b \in \fg,$ which accounts for the  sign difference between (\ref{thm4eq}) and (\ref{slinp}).}

\begin{equation}\labell{slinp}\langle (a,b), (a',b')\rangle_{\fg \oplus \fg} = {\rm vol}(M)( \langle a,b'\rangle_\fg + 
\langle b, a'\rangle_\fg).\end{equation}

Note that, for any $(f,g) \in F_1^F(M),$ 

\begin{equation}\labell{oddness} {\rm tr} \int_M \star g \wedge L f = - {\rm tr} \int_M \star f \wedge L g,
\end{equation}

\noindent so that to obtain the split linear pairing on $\fg \oplus \fg,$ the adjoint in $\cF^F_1(M)$ must include an additional negative sign on factors of $c_2.$   In other words we write for $(f,g) \in \cF^F_1(M)$

\begin{equation}\labell{defc1} f^c = f, g^c = -g,\end{equation}

\noindent or

\begin{equation}\labell{defc2} (f,g)^c = (f, -g),\end{equation}

\noindent and then

\begin{equation}\labell{evenness} {\rm tr} \int_M \star g^c \wedge L f =  {\rm tr} \int_M \star f^c \wedge L g ,
\end{equation}

\noindent so that we that the following restatement of Theorem \ref{rp1pf}:

\begin{Proposition}\labell{evennessp}
Let $\tilde{\Sigma}$ be a compact connected oriented two-dimensional submanifold of $M,$ and suppose $M - \tilde{\Sigma}$ consists of two components $U,V.$  Let $f \in \Omega^0(U,\fg)^\perp$ and $g \in {\rm ker} (d^*_2) \cap \Omega^2_c(V,\fg).$  Then 
\begin{equation}\labell{evenness1} {\rm tr} \int_M \star g^c \wedge L f =  {\rm tr} \int_M \star f^c \wedge L g = \langle \pi_{\tilde{\Sigma}}^F(f), \pi_{\tilde{\Sigma}}^F(g) \rangle_{\fg \oplus \fg}.
\end{equation}
 \end{Proposition}

The split linear pairing on $\fg \oplus \fg$ is not positive definite, but it is closely related to a positive definite pairing.  Let $J: \fg \oplus \fg \to \fg \oplus \fg$ be defined by

$$J(a,b) = (b,a).$$

Then the positive definite inner product on $\fg \oplus \fg$ given by

$$\langle (a,b) ,(a^\prime, b^\prime) \rangle_+ ={\rm vol}(M)( \langle a ,a^\prime \rangle + \langle b, b^\prime \rangle)$$

\noindent is related to the split linear pairing by

$$\langle \alpha, \beta \rangle = \langle \alpha ,J \beta \rangle_+$$

\noindent where $\alpha, \beta \in \fg \oplus \fg.$

To lift the positive pairing to  $F_1^F(M),$ we lift the map $J$ to a map $${\cdot}^t : F^F_1(M) \to F^F_1(M)$$ \noindent given by

$$(f,g)^t = (Lg, -\star d f).$$

Note that for $(f,g) \in F_1^F(M),$

\begin{equation} \labell{wicktinv}
{\rm tr}\int_M \star f \wedge L g = {\rm tr}\int_M \star f^t \wedge L g^t
\end{equation}

Then Theorem \ref{rp1pf} can be equivalently stated as

\begin{Proposition}\labell{rp1pfpos}
 Let $U,V$ be the components of $M - \tilde{\Sigma}.$  Let $f\in \Omega^0(U,\fg)^\perp,$ $\tilde{f}\in \Omega^0(V,\fg)^\perp$ and $g\in {\rm ker} (d^*_2) \cap \Omega^2_c(U,\fg),$ $\tilde{g}\in {\rm ker} (d^*_2) \cap \Omega^2_c(V,\fg).$

Then 

$${\rm tr}\int_M \star \tilde{f}^t  \wedge L f = - {\rm vol}(M) {\rm tr} c_f c_{\tilde{f}}  $$

and 

$${\rm tr}\int_M \star \tilde{g}^t  \wedge L g= - {\rm vol}(M) {\rm tr} c_g c_{\tilde{g}} . $$

\end{Proposition}
Alternatively, under the isomorphism 
$$ H^0(\tilde{\Sigma},\fg) \oplus H^2(\tilde{\Sigma},\fg) \simeq \fg \oplus \fg,$$
the split linear form is (up to a factor of ${\rm vol}(M)$) the bilinear form on $ H^0(\tilde{\Sigma},\fg) \oplus H^2(\tilde{\Sigma},\fg) $ arising from the intersection pairing of $H^0(\tilde{\Sigma},\fg)$ with $H^2(\tilde{\Sigma},\fg)$ and the inner product on $\fg.$ And we can consider the map $\pi_{\tilde{\Sigma}}^F$ as a map taking values in the one particle
Fermionic space $H^0(\tilde{\Sigma},\fg) \oplus H^2(\tilde{\Sigma},\fg).$  Note that this identification depends on an orientation of $\Sigma,$ and hence the reflection $R$ will introduce an additional sign in
the two form $c_2,$ which we have accounted for in the negative sign in the adjoint $(\cdot)^c$ mentioned above. 

Now consider the situation of Theorem \ref{rp1pf} in the case where the splitting of $M$ is given by $\Sigma.$   In the absence of Condition B, there is a subtlety about the definition of $R^*: \Omega^0(M_\pm,\fg)^\perp \to \Omega^0(M_\mp,\fg)^\perp,$ since elements of  $\Omega^0(M_\pm,\fg)^\perp$ are not compactly supported on $M_\pm $  ; instead they are constant outside of $M_\pm .$  However, if $f \in \Omega^0(M_\pm,\fg)^\perp,$ $f-c_f$ is compactly supported on $M_\pm.$  We define the map $R^*: \Omega^0(M_\pm,\fg)^\perp \to \Omega^0(M_\mp,\fg)^\perp$ for $ f \in \Omega^0(M_\pm,\fg)^\perp $ by

$$R^*f = R^*(f-c_f) + c_f.$$

 Then $R^*f \in \Omega^0(M_\mp,\fg)^\perp,$ and 
 \begin{equation}\labell{cfs2} c_{R^*f} = c_f.\end{equation} 
 
 \noindent Also,
 
 \begin{equation}\labell{cfs3}\star d(R^*f )= R^* (\star df).\end{equation}
 
In the absence of Condition A, we must restrict to forms constant outside of $M_\pm^\delta,$ and then the same procedure goes through.
 
We then have the following result:

\begin{corollary}\labell{rinvf1}    Let $f \in \Omega^0(M_-^\delta,\fg)^\perp$ and $g \in {\rm ker} (d^*_2) \cap \Omega^2_c(M_+^\delta,\fg).$  Then

\begin{equation}  {\rm tr}\int_M \star R^* f \wedge L R^*g = {\rm tr} \int_M \star   f\wedge L g.\end{equation}

In particular, $$c_{R^*g} = c_g.$$ 
(Where Condition A holds, we may replace $M_+^\delta$ by  $M_+$   and $M_-^\delta$ by  $M_-$ )
\end{corollary}
 \begin{pf} If Condition B is satisfied, this follows by $R-$invariance of the metric.  Otherwise, still, by Theorem \ref{rp1pf} and (\ref{cfs2})-(\ref{cfs3}), it suffices to prove that 

$${\rm tr}\int_M \star f \wedge L g  = {\rm tr}\int_{M} \star f \wedge R^*L R^* g.$$

 \noindent But $${\rm tr}\int_M \star f \wedge Lg  = {\rm tr}\int_M d L  f \wedge Lg= - {\rm tr}\int_M \star Lf \wedge g$$

 \noindent and, by a similar integration by parts, 

$${\rm tr}\int_M \star f \wedge R^*L R^* g   = -  {\rm tr}\int_M  Lf \wedge d R^*L R^* g  = - {\rm tr} \int_M \star Lf \wedge g.$$

 \end{pf}
A more subtle
reasoning is needed for the next result, which is an analog of Lemma \ref{rinv}.

\begin{Lemma}\labell{rinvf}Let $f \in \Omega^0(M_+,\fg)^\perp$ and $g \in {\rm ker} (d^*_2) \cap \Omega^2_c(M_+,\fg).$ Then

\begin{equation}\labell{rinvfe}{\rm tr}\int_M \star g \wedge L f= {\rm tr} \int_M \star R^*g \wedge LR^*f.\end{equation}

(In the absence of Condition A, replace $M_+$ by $M_+^\delta$ and $M_-$ by $M_-^\delta.$)\end{Lemma}

\begin{pf}  If Condition B is satisfied, this follows by $R-$invariance of the metric.

In the case of Condition A, we may not have a globally defined involution $R,$ but we can argue as in the proof of Lemma \ref{rinv}.

First, we compute for the right hand side of (\ref{rinvfe})\footnote{By a slight abuse of notation, we denote by $R^*LR^*f \in \Omega^2(M_+, \fg)$ the pullback $j^*((LR^*f)|_{M_-})$ of the form $(LR^*f)|_{M_-} \in \Omega^2(M_-, \fg)$  by the map $j:M_+\to M_-$ of (\ref{defj}).  The form $R^*LR^*f $ is not compactly supported, but its wedge product with the compactly supported form $\star g$ may be integrated on $M_+.$} 

$${\rm tr} \int_M \star R^*g \wedge LR^*f = {\rm tr} \int_{M_-} \star R^*g \wedge LR^*f= {\rm tr} \int_{M_+} \star g \wedge R^*LR^*f.$$

Then the difference between the right and left sides of (\ref{rinvfe}) is given by

$${\rm tr}\int_M \star g \wedge L f-  {\rm tr} \int_M \star R^*g \wedge LR^*f= {\rm tr}\int_{M_+} \star g\wedge (L f- R^*LR^*f).$$

But

\begin{align}\labell{rinvfe2}\begin{split}{\rm tr}\int_{M_+} \star g \wedge (L f- R^*LR^*f) = {\rm tr}\int_{M_+} dLg\wedge (L f- R^*LR^*f) =\\ -{\rm tr}\int_{M_+} Lg\wedge d(L f- R^*LR^*f)+ {\rm tr}\int_{\Sigma} Lg\wedge (L f- R^*LR^*f).\end{split}\end{align}

But, as forms on $M_+,$ 

$$d(Lf - R^*LR^*f) = \star(f - R^*R^*f) =0,$$

\noindent so that

$${\rm tr}\int_{M_+} Lg\wedge d(L f- R^*LR^*f)=0.$$

\noindent On the other hand

$$Lg|_\Sigma = c_g$$

\noindent so

$$ {\rm tr}\int_{\Sigma} Lg\wedge (L f- R^*LR^*f)=  {\rm tr}~ c_g\int_{\Sigma}  (L f- R^*LR^*f).$$

\noindent But

$$\int_{\Sigma}  (L f- R^*LR^*f) = \int_{M_+}  (dL f- dR^*LR^*f)= \int_{M_+}  (\star f- \star R^* R^*f)=0.$$

In the absence of Condition A, the same result holds for forms compactly supported on the complement $M_+^\delta$ of a collar neighborhood of $\Sigma.$

\end{pf} 

\subsection{Reflection positivity and $\cF^F$}\labell{rpfsub}

We consider the dual of the algebra $F^F(M)$ as generated by fields $c_0(f)$, $c_2(g)$ where $f \in \Omega^0(M,\fg)^\perp$ and $g \in {\rm ker}(d^*_2) \subset \Omega^2(M,\fg).$  On this algebra of fields we have the formal Berezin integral appearing in (\ref{start}), which we can use to produce a functional on on $F^F(M)$; viz\footnote{In this section, the wedge product $\wedge$ denotes the product in the alternating algebra $\cF^F$ and its variants, not the wedge product of differential forms.}

$$ \Psi(  f_1 \wedge \dots \wedge f_m  \wedge g_1\wedge \dots  \wedge g_n ) \sim \frac1{Z_{ghost}} \int dc_0 dc_2 e^{-{\rm tr} \int_M c_0 dc_2} c_0(f_1)\dots c_0(f_m) c_2(g_1)\dots c_2(g_n) $$

\noindent where $Z_{ghost}$ is a formally infinite normalization constant.

To define $\Psi$ in concrete terms that avoid Berezin integrals, let $f_i \in \Omega^0(M,\fg)^\perp, $ $ i = 1,\dots, m$ and $g_j\in {\rm ker}(d^*_2) \subset \Omega^2(M,\fg),$ $j = 1, \dots, n,$ and define

$$ \Psi( f_m \wedge \dots \wedge f_1  \wedge g_1\wedge \dots  \wedge g_n ) = \frac1{m!}{\rm det} M ,$$

\noindent where the $m \times m$ matrix $M$ is given by $M = 0 $ if $m \neq n$ and by

$$ M_{ij} = \langle f_i, L g_j\rangle, i, j = 1, \dots m$$

\noindent if $m = n.$

Alternatively, since the operator $\Delta^{-s/2} L \Delta^{-s/2}$ is antisymmetric and Hilbert Schmidt on the $L^2$ completion $\overline{\cF^F_1} $ of $\cF^F_1$ for our chosen $s=4,$\footnote{For this property $s > \frac14$ would suffice.} and hence gives an element $L_s\in \bigwedge^2 \overline{\cF^F_1}, $ we may write 

\begin{equation}\labell{defpsi}\Psi ( f_m\wedge \dots \wedge f_1 \wedge g_1\wedge \dots \wedge g_n) = \langle \frac{1}{n!} \bigwedge{}^{n} L_s, \Delta^{s/2}f_m \wedge \dots \wedge  \Delta^{s/2}f_1  \wedge \Delta^{s/2} g_1\wedge \dots  \wedge \Delta^{s/2} g_n\rangle_{\wedge^*\overline{\cF^F_1 } },\end{equation}

\noindent where the inner product is the inner product induced on the alternating algebra by the $L_2$ inner product of $\fg$-valued forms;  again this inner product is zero unless $m = n.$  This
shows, as in (\ref{bosebdd}) that 

$$|\Psi(P)| < C_d ||P||_{\cF^F_d}$$

\noindent for $P \in \cF^F_d.$  Thus $\Psi$ extends to a bounded linear functional on $\cF^F_d$ for every $d$, and to a linear functional
on $\cF^F_P.$\footnote{In fact, $\Psi$ is a bounded linear functional, but we do not make any use of this.}

As in the Bosonic case, it will be convenient to work with $L$-Wick ordered polynomials.\footnote{Some comments about Wick ordering:  As in the Bosonic case this is most conveniently described first for Hilbert Schmidt operators.

Suppose $L$ is an antisymmetric Hilbert Schmidt operator on a Hilbert space $H.$  Then there exists $v \in \wedge^2 H \subset H \otimes H$ with

$$\langle v, x \otimes y\rangle_{\wedge^2 H}  = \langle x, L y\rangle$$

for all $x,y \in H.$

Suppose $P \in \wedge^*H$ is a polynomial and define

$$\Psi(P) = \langle e^{i_v} P,1\rangle_{\wedge^* H}$$

where $i_v$ is interior product in $\wedge^*H;$ that is, 

$$i_v(x \wedge y) =\langle v, x \wedge y\rangle_{\wedge^2 H}.$$

Write, for $P \in \wedge^* H,$

$$:P: = e^{-i_v} P.$$

Then if $P,Q \in \wedge^* H,$

\begin{equation}\label{eq2}\Psi(:P::Q:) = \langle P, (\sum_{k=0}^\infty  \frac1{k!}\wedge{}^k L )Q \rangle_{\wedge^* H}. \end{equation}

As in the Bosonic case, if $L$ is not Hilbert Schmidt, similar formulas hold as long as both sides of equation (\ref{eq2}) are well defined; these can be proved by cutting off $L$ to obtain a sequence of finite rank operators, and then taking the limit on both sides.  In our case a transfer of regularity argument can be used, involving the operators $L_s$ of equation (\ref{defpsi}).}

A compact way of defining these is by taking another copy $X$ of $\cF^F_1$ and defining

$$ : e^{\langle x, y\rangle} : = e^{- \frac12 \langle x, L x\rangle} e^{\langle x, y \rangle}$$

\noindent for any $x \in X,$ $y \in \cF^F_1$ (see \cite{f}, page 22); as in the Bosonic case, the Wick ordered polynomials can be extracted from this formula.

As in the previous section, suppose $\tilde{\Sigma}$ is a smooth compact connected submanifold of $M$ splitting $M$ into two components $U, V.$ We now extend
the definition of the map $\pi^F_{\tilde{\Sigma}}$ to a map $\pi^F_{\tilde{\Sigma}}: \cF^F_P(U) \to \bigwedge^*({\fg \oplus \fg})$ by 

$$\pi^F_{\tilde{\Sigma}} (:  f_1\wedge\dots \wedge f_n\wedge g_1\wedge\dots \wedge g_m:)
= \pi^F_{\tilde{\Sigma}} (f_1)\wedge \dots \wedge \pi^F_{\tilde{\Sigma}} (f_n)\wedge
\pi^F_{\tilde{\Sigma}} (g_1)\wedge \dots \wedge \pi^F_{\tilde{\Sigma}} (g_m)$$

\noindent and extending by linearity, and similarly for forms compactly supported in $V.$  This map is again
bounded on each $\cF^F_{ d}(U)$ and each $\cF^F_{ d}(V).$

Given a monomial $m,$ we extend the definition of the map $(\cdot)^c$ in (\ref{defc1}) by writing 

\begin{equation}\labell{defc} m^c = (-1)^{s_1 + s_2}m,\end{equation}

\noindent where $s_1 = \frac{|m| (|m|-1)}{2},$ $|m|$ is the degree of $m$, and $s_2$ is the number of factors of $c_2$ in $m.$  Intuitively, this monomial is obtained by reversing the order of the terms in $m$ and attaching a $-$ sign to each factor of $c_2.$ Explicitly, if 
$$ m = f_1 \wedge \dots \wedge f_k \wedge g_1 \wedge \dots \wedge g_l,$$

\noindent where $f_i, g_j$ are as above, then

$$ m^c = -g_l \wedge \dots \wedge -g_1 \wedge f_k \wedge \dots f_1.$$

Note that for any monomial $m$,

\begin{equation}\labell{wickc} :m^c: = :m:^c.\end{equation}

Likewise we extend the definition of the map ${\cdot}^t$ to monomials by defining $m^t,$ where $ m = f_1 \wedge \dots \wedge f_n \wedge g_1 \wedge \dots \wedge g_m,$ by

$$m^t = Lg_m \wedge \dots \wedge Lg_1 \wedge( -\star d f_n) \wedge \dots \wedge (-\star df_1).$$  Extending by linearity, the definitions of $\cdot^c$ and $\cdot^t$ extend to all polynomials in $\cF^F_P(M).$   

The following proposition is then an immediate consequence of the definition of $\Psi$ and Theorem \ref{rp1pf}.

\begin{Proposition}\labell{piisom} Let $p \in \cF^F_P(U)$ and $q \in \cF^F_P(V).$   Then 

$$\Psi(:p^c: \wedge :q:) = \langle \pi^F_{\tilde{\Sigma}}(:p:), \pi^F_{\tilde{\Sigma}}(:q:)\rangle_{\bigwedge^*(\fg \oplus \fg)}$$

\noindent and therefore by (\ref{wickc})

$$\Psi(p^c \wedge q) = \langle \pi^F_{\tilde{\Sigma}}(p), \pi^F_{\tilde{\Sigma}}(q)\rangle_{\bigwedge^*(\fg \oplus \fg)}$$\\

\noindent where the inner product on $\bigwedge^*(\fg \oplus \fg)$ is that given by the split linear form on $\fg \oplus \fg.$\\

In particular, if $\tilde{\Sigma} = \Sigma, U = M_+, V=M_-,$

$$\Psi((R^*p)^c \wedge p) = \langle \pi^F_{{\Sigma}}(p), \pi^F_{{\Sigma}}(p)\rangle_{\bigwedge^*(\fg \oplus \fg)}.$$

\end{Proposition}

Likewise we have

\begin{Proposition}\labell{rpfcor}

Let $p \in \cF^F_P(U)$ and $q \in \cF^F_P(V).$   Then

$$\Psi(:p^t:\wedge :q:) = \langle \pi^F_{\tilde{\Sigma}}(:p:), \pi^F_{\tilde{\Sigma}}(:q:)\rangle_{+}$$

\noindent where the metric $\langle\cdot,\cdot\rangle_+$ on $\bigwedge^*(\fg \oplus \fg)$ is the one given by the positive metric $\langle\cdot,\cdot\rangle_+$ on $\fg \oplus \fg.$
\end{Proposition}

Proposition \ref{rpfcor}, along with the reflection invariance of the Wick ordering (Corollary \ref{rinvf1} and Lemma \ref{rinvf}), hint at the possibility of a reflection positivity theorem for the fermions.  However, the Wick ordering $:\cdot:$ is not invariant under the transformation given by $(\cdot)^t$, and in fact the Fermions contain negative norm ("ghost") states.\footnote{For example, consider $p=  (Lg \otimes \xi) \wedge (g \otimes \xi)$ where $g \in {\rm ker} (d_2^*) \subset \Omega^2_c(M_+^\delta,\R)$ satisfies $c_g = 0$ and $\xi$ is a nonzero element of $\fg.$  Then  $\Psi(Rp^t p) = - ||\xi||^2_{\fg}||Lg||^2_2.$}    We will thus have to handle the Boson-Fermion interaction term by different methods than those used for the Boson interaction term, which is a self adjoint operator.  

In any event, we write
 $H^F(\Sigma) = \bigwedge^* (\fg \oplus \fg),$
with the positive definite metric $\langle \cdot , \cdot \rangle_+.$\footnote{As for the gauge fields, where the symmetric algebra is a subalgebra of a Weyl algebra, this alternating algebra is embedded in a Clifford algebra on two copies of $\fg \oplus \fg$ corresponding to ghost fields compactly supported on both sides of $\Sigma.$} 

Likewise if $\tilde{\Sigma}$ is any two dimensional submanifold of $M$ isotopic to $\Sigma,$ and $U,V $ are the components of $M - \tilde{\Sigma},$  we write $H^F(\tilde{\Sigma})$ for the Hilbert space associated to $\tilde{\Sigma}$ (which is of course isomorphic to $H^F(\Sigma)$) to make the relation to the geometry explicit.

Note that since  $H^F(\Sigma)$ is an alternating algebra on a finite dimensional vector space, it is equipped with a bounded, nonnegative, self-adjoint operator $N_F :  H^F(\Sigma) \to  H^F(\Sigma)$ given by polynomial degree.  Also, since  $H^F(\Sigma)$ is finite-dimensional, the two inner products on  $H^F(\Sigma)=\bigwedge^* (\fg \oplus \fg)$ are related by a (necessarily bounded) invertible linear operator $\cJ_F = \bigwedge^*J:   H^F(\Sigma)  \to   H^F(\Sigma).$  
 
 \begin{Remark}\labell{fscaling}As in the Bosonic case, scaling the quadratic term in the free Lagrangian by a factor of $\alpha$ corresponds to replacing the functional $\Psi$ by a functional $\Psi_\alpha$ obtained by multiplying the right hand side of (\ref{defpsi}) by a factor of $\alpha^{-n}.$  Equivalently, the inner products on $H^F(\Sigma)$ can be modified by a factor of $e^{-\log(\alpha) N_F}.$\end{Remark}

\subsection{The Hilbert space $H(\Sigma)$}We write

$$ H(\Sigma) = H^B(\Sigma) \otimes H^F(\Sigma).$$  We also write $$\pi_{\Sigma} = \pi_{\Sigma}^B\otimes \pi_{\Sigma}^F,$$ 

\noindent reiterating that the map $\pi_\Sigma$ is only defined on the dense subspace $\cF_{P,+} \subset \cF_+.$  Similarly $H(\tilde{\Sigma}), \pi_{\tilde{\Sigma}}$ for any compact connected smooth submanifold $\tilde{\Sigma}$ isotopic to $\Sigma.$ 

The Hilbert space  $H(\Sigma)$ is equipped with a densely defined, nonnegative, self-adjoint operator $N$  given by $N = N_B + N_F,$ as well as with a bounded operator $\cJ : H(\Sigma) \to  H (\Sigma)$ given by $\cJ= {\rm id}_{H^B(\Sigma)} \otimes \cJ_F.$  And as before, a scaling of both quadratic terms in the free Lagrangian by a factor of $\alpha$ is equivalent to a modification of the metric on the Hilbert space $H(\Sigma)$ by the positive bounded self-adjoint operator  $e^{-\log(\alpha) N}.$

We extend the action of the conjugation $(\cdot)^c$ to $\cF_P(M)$ by defining its action on Bosonic fields to be trivial.  Likewise define the map $R^*: \cF_P(M_+ ) \to \cF_P(M_- )$  as the tensor product of the maps (both denoted $R^*$) $R^*:\cF_P^B(M_+ )\to \cF_P^B(M_- )$ and $R^*:\cF_P^F(M_+ )\to \cF_P^F(M_-),$ with the obvious replacement of $M_\pm$ with $M_\pm^\delta$ if Condition A is not satisfied.  Then the following Theorem summarizes the relation between the functionals $\Phi$ and $\Psi$ (which correspond to expectations in the formal functional integral arising from the quadratic part of the Chern Simons Lagrangian) and the Hilbert space $H(\Sigma).$\\
 
\begin{Theorem}\labell{ospcombined}

Let $p, q \in \cF_{P,+}.$   Then

$$(\Phi \otimes \Psi) ( (R^*p)^c q ) = \langle \pi_\Sigma(p), \cJ \pi_\Sigma(q)\rangle.$$

\end{Theorem}

\section{The interacting Theory}\labell{intsec}
 
So far we have constructed a Hilbert space $H(\Sigma) = H^B(\Sigma) \otimes H^F(\Sigma)$ 
with a positive definite inner product in which correlations of polynomial
functions are as would be expected from the quadratic part of the formal
Chern-Simons functional integral with action

$$S_F(A,c_0,c_2) = \frac{1}{2}  {\rm tr} \int_{M} A d A - 2 c_0 dc_2.$$

We now wish to include the interaction term\footnote{We focus in this introduction on the case where Condition A holds.}

$$S_I(A,c_0,c_2) =  {\rm tr} \int_{M} A^3 - 6 A c_0 c_2 .$$

These interaction terms are local, and are morally of the form

$$ S_I = \Xi_B^+ + R^*\Xi_B^+ + \Xi_{BF}^+ + R^* (\Xi_{BF}^+)^c,$$

\noindent where 

\begin{equation}\labell{dxb} \Xi_B^+(A) =  {\rm tr} \int_{M_+} A^3\end{equation}

and

\begin{equation}\labell{dxf} \Xi_{BF}^+ (A,c_0,c_2) =- 6 {\rm tr} \int_{M_+}  A c_0 c_2\end{equation}

\noindent are local interaction terms that have the appearance of elements of $\cF_{3,+},$\footnote{Note that $\Xi_B^+$ and $\Xi_{BF}^+$ are Wick ordered polynomials.} and hence should yield polynomials of degree $3$ in $H(\Sigma).$

However, a closer study of the terms $\Xi_B^+ $ and $\Xi_{BF}^+$ shows that in order to interpret them as elements of $H(\Sigma),$ some choices are necessary.  This is due to the fact that elements of $\cF_{P,+}$ consist of local polynomials built from linear factors arising from differential forms compactly supported in\footnote{In the case of the zero form, with {\em derivative} compactly supported in $M_+$} $M_+$ and satisfying the gauge fixing conditions $d^*(\cdot) = 0,$ while the integrals over $M_+$ appearing in (\ref{dxb}) and (\ref{dxf}) are not of this form. To see this, choose an orthonormal basis $\{e_\alpha\}$  for $\fg$ and three smooth one forms $f,g,h \in \Omega^1_c(M_+)$ satisfying the gauge condition.  We would expect, morally, 

\begin{equation}\labell{defint} \Phi(A(R^*f\otimes e_\alpha) A(R^*g \otimes e_\beta) A(R^*h \otimes e_\gamma) \Xi_B^+)  = - f_{\alpha\beta\gamma} \int_{M_+} LR^*f \wedge L R^*g \wedge L R^*h.\end{equation}

\noindent where $f_{\alpha\beta\gamma}= -{\rm tr} (e_\alpha[e_\beta,e_\gamma]).$  Now the integral

$$ \int_{M_+} LR^*f \wedge L R^*g \wedge L R^*h $$

\noindent does depend only on the boundary values $Lf|_\Sigma, Lg|_\Sigma, Lh|_\Sigma \in Z^1(\Sigma);$ but not only on the cohomology classes $[Lf|_\Sigma], [Lg|_\Sigma], [Lh|_\Sigma] \in H^1(\Sigma).$   Similar problems arise for $\Xi_{BF}^+.$  So it is not possible to interpret the interaction terms as elements of $H(\Sigma)$ without making choices, specifically of an orthonormal basis $\{ b_i\}$ for $\Lambda \subset H^1(\Sigma)$ and of one forms $\psi_i$with $[L{\psi_i}|_\Sigma]= b_i,$ along with a similar choice for the Fermions.  Once such a choice is made, equation (\ref{defint}) defines an element $\xi_B^+ \in H_B(\Sigma)$ of degree 3.  This gives a densely defined multiplication operator $\xi_B^+ \cdot,$ and a symmetric operator $\xi_B^+ \cdot + (\xi_B^+ \cdot)^*,$ which has a self adjoint extension $O_B$ and gives rise to a one parameter subgroup $e^{i \lambda O_B}$ on $H_B(\Sigma) \otimes \C.$

Similarly, the Fermion interaction term $\Xi_{BF}^+$ gives rise, after choices, to an element $\xi_{BF}^+ \in H(\Sigma)$ which is linear in the gauge field and quadratic in the Fermions.  The interaction term does not correspond to the sum of this element and its adjoint in the positive metric on $H (\Sigma),$ but rather to the sum of this element and its adjoint in the indefinite inner product on $H(\Sigma)$ arising from the split linear metric on $\fg \oplus \fg.$  It therefore does not give a self adjoint operator on a Hilbert space, and does not give rise to a one parameter subgroup of unitary operators.  However, we can consider the operator

$$O_{BF} = \xi_{BF}^+ \cdot + ( \xi_{BF}^+ \cdot )^\dag$$

\noindent where $(\cdot)^\dag = \cJ^{-1} (\cdot)^* \cJ$ denotes the adjoint in the indefinite inner product on $H(\Sigma)$ arising from the split linear metric on $\fg \oplus \fg.$  The operator is defined on the dense subset of $H(\Sigma)$ consisting of elements of finite degree.  Since $\xi_{BF}$ is linear in the gauge field, and since the ghost Hilbert space is finite dimensional, so that the ghost field operators are bounded, we have a bound

$$O_{BF}^* O_{BF} \leq K_1 N_B + K_2$$

\noindent for some constants $K_1,K_2.$  We can therefore exponentiate $O_{BF}$ using the explicit power series to obtain a family 

$$ e^{-\epsilon N} e^{i\lambda O_{BF}}$$

\noindent for every $\lambda \in \R$ and $\epsilon >0$ which are bounded for every $\epsilon,$ and extend to all of $H(\Sigma).$\footnote{A similar method for handling Bose-Fermi interactions is in \cite{os2}.}  We can then combine the operators $e^{i \lambda O_B}$ and $ e^{-\epsilon N} e^{i\lambda O_{BF}},$ and take a weak limit inspired by the Trotter product formula, to obtain what {\em morally} should be the partition function

$$ \langle \Omega ,\cJ e^{-\epsilon N + i \lambda (O_B + O_{BF})}\Omega\rangle,$$

\noindent where $\Omega = \pi_\Sigma(1).$  Since we have seen (see Remark \ref{bscaling} and Remark \ref{fscaling}) that a factor of $e^{-\epsilon N}$ amounts to a renormalization of the free Lagrangian, this is a reasonable definition of the partition function.\footnote{Our actual construction is slightly different.  See Section \ref{finalpf}.}

In the remainder of this section, we focus on the case where Condition A is satisfied.  If this is not the case, an additional weak limit has to be taken as $\delta \to 0.$

\subsection{The Boson interaction term}\labell{bintterm}
We now define the Bosonic interaction term.  Choose an orthonormal basis $\{e_\alpha\}$ for $\fg$ and write $f_{\alpha\beta\gamma} = -{\rm tr} (e_\alpha, [e_\beta,e_\gamma]).$
Choose a basis $b_1,\dots,b_g$ for $\Lambda \subset H^1(\Sigma), $ and, using Proposition \ref{h1f1}, choose one forms 
$\phi_1,\dots,\phi_g \in F_1^B(M_+)$ such that $$[L\phi_i\vert_\Sigma] = b_i.$$  

We define the Bosonic interaction element 
$$\xi_B^+ \in {\rm Sym}^3(\Lambda) \subset H^B(\Sigma)$$ 

\noindent as the unique element of the finite dimensional vector space ${\rm Sym}^3(\Lambda)$  satisfying  

$$\langle\xi_B^+, (b_i \otimes e_\alpha) \otimes (b_j \otimes e_\beta) \otimes (b_k \otimes e_\gamma) \rangle = -f_{\alpha\beta\gamma} \int_{M_+} LR^*\phi_i \wedge LR^*\phi_j \wedge LR^* \phi_k$$

\noindent for all $\alpha,\beta,\gamma, i, j, k.$  

Then $ \xi_B^+$ gives a densely defined multiplication operator $ \xi_B^+ \cdot$ on $H^B(\Sigma),$ with domain $\cD = {\rm Sym}^* \Lambda \subset H^B(\Sigma) $ given by the elements of finite degree.  Since the domain of the adjoint $ (\xi_B^+ \cdot)^* $ contains $\cD,$  the sum 

$$ \xi_B^+ \cdot + (\xi_B^+ \cdot)^*$$

\noindent is a densely defined symmetric operator on the real Hilbert space $H^B(\Sigma), $ and hence has a (possibly non-unique) self-adjoint extension $O_B$ densely defined on $H^B(\Sigma) \otimes \C.$\footnote{This is because a densely defined symmetric operator on a real Hilbert space has equal deficiency indices as on operator on the complexified Hilbert space.  See e.g. \cite{lax}, p. 402, or \cite{rs} p. 143.}  We may then exponentiate $O_B$ to obtain a one parameter subgroup 

$$ e^{i\lambda O_B} : H^B(\Sigma)\otimes \C \to H^B(\Sigma) \otimes \C$$

\noindent of unitary operators.

By a slight abuse of notation we use the notation $e^{i\lambda O_B}$ also for the one parameter subgroup of unitary operators acting as $e^{i\lambda O_B} \otimes {\rm id}$ on $H^B(\Sigma) \otimes H^F(\Sigma) \otimes \C = H(\Sigma)\otimes \C.$

\begin{Remark}\labell{bhilbr} The fact that the interaction term $\Xi_B^+$ is not naturally an element of $H^B(\Sigma)$ raises the issue of whether a larger Hilbert space, big enough to contain $\Xi_B^+,$ can be constructed from polynomials in the gauge fields built from $\fg-$valued one forms compactly supported on $M_+$ but not satisfying the gauge conditions.  Given $f,g \in \Omega^1_c(M_+,\fg)$, we may compute, as in Lemma \ref{L to coho},

\begin{equation}\labell{bigger} {\rm tr}  \int_M \star \pi(f) \wedge R^* L \pi(g)\end{equation}

\noindent where

$$\pi  = d^* d \Delta^{-1} $$

\noindent is the projection onto one forms satisfying the gauge condition.  A computation along the lines of that in Lemma \ref{L to coho} again reduces the integral in (\ref{bigger}) to an integral over $\Sigma,$ involving now three terms:\footnote{In this equation the $\star$ operator and the expression $d^*$ refer to the Hodge star operator on $M,$ not on $\Sigma.$} 

\begin{equation}\labell{bigger1}{\rm tr}  \int_M \star \pi(f) \wedge R^* L \pi(g) ={\rm tr} \int_\Sigma \star d \Delta^{-1} f \wedge \star d \Delta^{-1} R^*g + d^* \Delta^{-1} f \wedge d \Delta^{-1} R^*g  +   d \Delta^{-1} f \wedge d^* \Delta^{-1} R^*g
\end{equation}
The first term pairs a one form (now not necessarily closed) arising from $f$ with a similar one form arising from $g,$ as in Lemma \ref{L to coho}, and the two additional terms pair zero- and two-forms arising from $f$ with similar forms arising from $g.$  It would be interesting to see if any form of reflection positivity holds for the pairing given by (\ref{bigger1}), and to try to identify the resulting Hilbert space.  The Hilbert space $H^B(\Sigma)$ would then be a subspace of the larger Hilbert space hinted at by (\ref{bigger1}), and the element $\xi_B^+ \in H^B(\Sigma)$ would be a projection of the element $\Xi_B^+$ to this subspace. It is also conceivable that the effects of the terms corresponding to the pairing of zero-forms with two-forms cancel in some way with the contributions from the ghost fields, so that we effectively end up again with something like $H^B(\Sigma).$  
\end{Remark}

\subsection{The Fermion interaction term}\label{fintterm}

In order to define the Fermion interaction term, we must supplement the choices made in the previous section by choosing also a function $f \in \Omega^0(M_+,\fg)^\perp$ with $c_f = 1.$    We define the Fermion interaction element as the unique element $\xi_{BF}^+ \in \Lambda \otimes H^F(\Sigma) \subset H(\Sigma)$ satisfying  

$$\langle  \xi_{BF}^+, \cJ \left((b_i \otimes e_\alpha) \otimes (e_\beta \oplus 0) \wedge (0 \oplus e_\gamma)\right)\rangle = 6  f_{\alpha\beta\gamma} \int_{M_+} LR^*\phi_i \wedge LR^*f $$

\noindent for all $\alpha,\beta,\gamma, i,$ where we have identified the one particle subspace in $H^F(\Sigma)$ with $\fg \oplus \fg.$

As before, the element  $ \xi_{BF}^+$ gives a multiplication operator $ \xi_{BF}^+ \cdot$ densely defined on $\cD \otimes H^F(\Sigma).$  A direct computation shows that the domain of the adjoint operator $ (\xi_{BF}^+ \cdot)^*$ contains $\cD \otimes H^F(\Sigma).$

Then the Fermion interaction term is given by the operator

$$O_{BF} =  \xi_{BF}^+ \cdot + ( \xi_{BF}^+\cdot)^\dag,$$

\noindent where the notation $\cdot^\dag$ denotes the adjoint in the bilinear form on $H(\Sigma)$ given by the usual metric on $H^B(\Sigma)$ and the split linear form on $H^F(\Sigma);$  in terms of the Hilbert space metric, we have

$$( \xi_{BF}^+\cdot)^\dag = \cJ^{-1} ( \xi_{BF}^+\cdot)^*  \cJ$$

\noindent where $\cJ$ is the bounded linear operator on $H (\Sigma)$ defined in Section \ref{rpfsub} arising from the linear transformation on $H^F(\Sigma)$ relating the split linear form to the metric.  Since the operator $\cJ$ acts as the identity on $H^B(\Sigma),$ it preserves the domain $\cD \otimes H^F(\Sigma)$, so that the operators $( \xi_{BF}^+\cdot)^\dag$ and  $O_{BF}$ are also defined on the dense domain $\cD \otimes H^F(\Sigma).$

Now the operator $O_{BF}$ is not symmetric, so it cannot be extended to a self-adjoint operator and exponentiated to a one parameter subgroup of bounded operators.  However, $\xi_{BF}^+$ is linear in the gauge fields, so we have the inequality

\begin{equation}\labell{obfest}  O_{BF}^* O_{BF} \leq K_1 N_B + K_2 \end{equation}

\noindent for some constants $K_1,K_2\geq 0.$

We may therefore define, on $\cD \otimes H^F(\Sigma)\otimes \C$, and for any $\lambda \in \C,$ the operator 

$$e^{-\epsilon N_B} e^{ \lambda O_{BF}} := \sum_{k=0}^\infty \frac{1}{k!} \lambda^k e^{-\epsilon N_B} (O_{BF})^k$$

\noindent which, by (\ref{obfest}), is bounded for every $\epsilon > 0,$ and hence gives an operator on $H(\Sigma)\otimes \C;$

\noindent and we have the bound

\begin{equation}\labell{obfbd}|| e^{-\epsilon N_B/n} e^{i \lambda O_{BF}/n} || \leq K^{\frac{1}{n}} \end{equation}

\noindent for some constant $K \geq 0$ (which depends on $\epsilon$ and $\lambda$ but not on $n$) for any $n > 0.$

\subsection{The Partition function}\labell{finalpf}

In this section we combine the results of Section  \ref{bintterm} and Section \ref{fintterm} to define the partition function of our theory.

Let $\lambda, \epsilon \in \R$ and, for any $n > 0,$ let

\begin{equation}\labell{partdef} 
Z_n(\epsilon,\lambda) = \langle \Omega , \cJ ( e^{-\epsilon N/n} e^{i \lambda O_{BF}/n} e^{i\lambda O_B/n})^n \Omega \rangle
\end{equation}

\noindent where $\Omega$ is the identity element $\Omega = 1 \in {\rm Sym}^* \Lambda \otimes \bigwedge^* (\fg \oplus \fg).$ 

By the estimate (\ref{obfbd}), the sequence $Z_n$ is bounded, so that   
\begin{Theorem}\labell{pfdef}  The weak limit 

$$ Z (\epsilon,\lambda)= {\rm wk} \lim_{n \to \infty} Z_n(\epsilon,\lambda).$$

\noindent exists.  

\end{Theorem}

The function $ Z (\epsilon,\lambda)$ is the partition function of the quantum field theory.\footnote{This method of taking limits through subsequences appears in constructive quantum field theory in Glimm-Jaffe's construction of the infinite
volume limit in two dimensions in \cite{gjacta}.  In that
case, also, establishing uniqueness of the limit required further developments \cite{GJ}.}  Note that the theory is finite and does not require renormalization, as may be expected from perturbation theory \cite{as}.

In view of the Trotter product formula, and of the fact that inserting a factor of $e^{-\epsilon N}$ in the inner product amounts to a scaling of the free Lagrangian (see Remark \ref{bscaling} and Remark \ref{fscaling}), this seems like a reasonable definition.\footnote{It is also possible to contemplate a more elaborate construction, motivated by a time ordered exponential, obtained by cutting the manifold $M_+$ and its mirror image into $N$ slices, and viewing the corresponding interaction terms as giving a map between the Hilbert spaces corresponding to the boundaries of the slices.  We would then compose the exponentials of these terms and take another weak limit as $N \to \infty.$}  But several questions come to mind.

\begin{Question} \labell{conj1}
Is the weak limit unique?
\end{Question}

\begin{Question} \labell{conj2}
Assuming Conjecture \ref{conj1}, is this limit independent of choices?\footnote{A speculation:  The classical object corresponding to the sum of $\xi_B^+$ and its adjoint vanishes morally due to the vanishing of the cup product $\bigwedge^3 H^1(M) \to H^3(M).$  Another aspect of this is the cancellation of the boundary terms in (\ref{defint}) between $\Xi_B^+$ and its reflection.  It might be that the quantization $O_B,$ which is nonvanishing, may give rise to a partition function independent of choices as a type of anomalous quantum correction to the vanishing of the classical invariant.}
\end{Question}

\begin{Question} \labell{conj3}
Assuming Conjecture \ref{conj1} and Conjecture \ref{conj2}, is this limit a topological invariant of $M?$  How is it related to the invariants of Witten, Reshetikhin, and Turaev?
\end{Question}

It may be that a positive answer to these questions can be expected if the coupling constant is quantized.

More generally, answering these questions may require ideas beyond reflection positivity.  But reflection positivity does provide a context in which those questions can be formulated precisely.

\begin{Remark}\labell{moralpf}
Recalling the discussion in Section \ref{Kto0}, we see that morally, the partition function should be a putative expression of the form

\begin{equation}\labell{formalcorrectpf}
\langle \pi_\Sigma (e^{-i \lambda( \Xi_B^+ + \Xi_{BF}^+)} ) ,  \cJ \pi_\Sigma (e^{i \lambda( \Xi_B^+ + \Xi_{BF}^+)})  \rangle_{H(\Sigma)\otimes \C}.\end{equation}

\noindent  Such an expression would be expected if we were to construct a topological field theory, which assigns an element of the Hilbert space associated to a two-manifold to any bounding three-manifold.  And generalizations of (\ref{formalcorrectpf}) would then be available for any compact three manifold, not only for manifolds with reflection. However, as we noted in Section \ref{Kto0}, and even leaving aside the fact that $\Xi_B^+$ and $\Xi_{BF}^+$ are not quite elements of $\cF_+,$ we do not know how to construct a map $\cF_+^B \to H^B(\Sigma),$ and hence we do not know how to make sense of the expression $\pi_\Sigma (e^{i \lambda( \Xi_B^+ + \Xi_{BF}^+)})$ unless we replace the operator $L$ appearing in all our definitions of $\Psi, \Phi,$ etc. by the operator $K I + L$ where $K > ||L||.$   The expression (\ref{formalcorrectpf}) therefore makes
sense only for $K$ large, not for the geometrically relevant case $K = 0.$  

The expression (\ref{partdef}), while well defined for the geometrically relevant covariance $L,$ is only available for manifolds with reflection.

This raises the issue of how an exponential of the form (\ref{partdef}) is morally related to a putative partition function of the form (\ref{formalcorrectpf}).  One way to look at  $\xi_B^+ + \xi_{BF}^+$ as an element of the Hilbert space of a three manifold with one boundary $\Sigma$ and one "boundary" consisting of a point, to which topological quantum field theory would naturally assign the vacuum.  The composition in topological quantum field theory of two such elements reduces essentially to multiplication, and so the exponentials in the morally correct (\ref{formalcorrectpf}) may be expected to be expressed in the mathematically well-defined (\ref{partdef}).

One way to address these issues is via Questions \ref{conj1}-\ref{conj3}. \end{Remark}

\section{Concluding remarks and speculations}\labell{conclu}

We conclude with some remarks and speculations about Chern-Simons functional integrals.

We begin with a general comment about the potential role of Chern-Simons functional integrals.  First, and quite apart from topological quantum field theory, the question of providing a mathematical interpretation of functional integrals is of independent interest, and a longstanding problem in Mathematical Physics.  Second, the construction we give indicates a possible close relationship between the topological quantum field theories associated with abelian Lie groups, which are related to the quadratic term in the Chern Simons Lagrangian, and the topological quantum field theories associated to nonabelian groups.   It would appear from the functional integral that nonabelian theories arise in some version of the abelian theory.  This phenomenon may be interesting to study from a topological point of view.  Finally, if functional integrals give an alternative construction of topological quantum field theories, and these were related to those constructed by topological methods, such a relation between topology and analysis could be seen as an infinite dimensional nonlinear analog of the de Rham theorem, or perhaps of the Atiyah-Singer index theorem. 
 
\subsection{A  finite dimensional expression that may be related to the Chern Simons path integral}\labell{fpisec}

The Hilbert Space $H^B(\Sigma)$ is the completion of a space of polynomials on a finite dimensional vector space $\Lambda \subset H^1(\Sigma,\fg).$  The partition
function and expectations of polynomials, which are morally given by integration on an infinite dimensional space of connections, should then also
be expressible as path integrals over this finite dimensional space.  The Fermion Hilbert space $H^F(\Sigma)$ is finite dimensional, and the action is
quadratic in the ghost variables, so the Fermi integral can be performed explicitly to add a determinant to this integral.  One may speculate that since this is a topological quantum field theory, the path integral may be related to a finite dimensional integral.
Where $M$ is Heegaard split along $\Sigma,$ we may consider an integral over the
finite dimensional vector space $\R^{2g}\otimes \fg$ roughly of the form

\begin{equation}\labell{findimpi}
\int_{\Lambda } (\prod_{a,\alpha} dt_a^\alpha) e^{- {} \langle t, (J^{-1} \otimes I) t\rangle - i \lambda \sum_{\alpha,\beta,\gamma}\sum_{a,b,c} t_a^\alpha t_b^\beta t_c^\gamma f_{\alpha\beta\gamma}N_{abc} } 
{\rm det} ({} K\otimes I + 6 i \lambda \sum_{a, \alpha} t_a^\alpha B_a \otimes F_\alpha)\end{equation}

\noindent where the elements of the matrices $J, K,B_a$, and $F_\alpha$ and the
tensor $N_{abc}$ are constants depending on the manifold $M$ and
the Heegaard splitting.  For example, the constants $N_{a,b,c}$ may be computed for elements $a,b,c$ of the cohomology of $\Sigma$ by extending closed forms representing them  to forms $A,B,C$ on $M$ which are closed on $M_-$ and computing the triple intersection ${\rm tr} \int_M \star d A \wedge \star d B \wedge \star d C,$ and then adding the result to a similar computation performed by exchanging $M_+$ with $M_-$ and replacing $a,b,c$ with $S^*a, S^*b,S^*b.$ And the bilinear form $J$ arises from the intersection form on $\Sigma$ and the diffeomorphism $S.$ Then the positivity of the matrix $J$   should imply that this integral is a matrix Airy
function of the form studied in \cite{konts}.   (The determinant appearing in the integrand is a polynomial in the $t_a^\alpha,$ therefore integrable.)   

Alternatively there may be as we mentioned a formulation in terms of an integral over a space of paths on $\Lambda.$\footnote{The fact that the cup product map $\bigwedge^3 H^1(M) \to H^3(M)$ vanishes may mean that an integral on the space of paths, with a path of connections on $\Sigma$ ending at the trivial connection giving rise to a connection on $M_+,$ is a more plausible construction.  Note that the deformation of algebras of functions on a Poisson manifold referred to below also amounts to a functional integral over paths \cite{cf}.}

The role of the group $G$ is quite minimal in this finite dimensional 
path integral; in fact it appears that all that matters is the abelian
path integral, which gives the matrix $J,$ and some kind of lie algebra
to make the alternating tensor $N$ into a reasonable integrand.

In a more speculative vein, the perturbation of a quadratic integral on a symplectic manifold by a cubic term may hint at a deformation of the quantization of a ${\rm dim}(G)$-fold product of the Jacobian of a Riemann surface by a term determined by a bounding three manifold.  This deformation involves the construction of a cubic term on $H^1(\Sigma,\fg)$ using the Hodge star operator on a bounding three manifold and the structure constants of the Lie algebra $\fg.$  This is reminiscent of the deformation quantization of \cite{konts2}.\footnote{Perhaps the tensors $\epsilon_{ijk}$ and $f_{\alpha\beta\gamma}$ combine to give a kind of even analog of a Poisson structure on $\cA.$  A similar appearance in Chern Simons gauge theory of an even analog of a symplectic form arising from the quadratic term in Chern-Simons gauge theory appears in \cite{w1}.}  In other contexts, the role of the Hodge star operator and of cubic deformations of algebras in finding manifold invariants beyond cohomology has been emphasized by D. Sullivan. 

\subsection{Relation to topological quantum field theory} 
The first aspect of a construction of topological quantum field theory would be to show that the limit partition function is a topological invariant.  Beyond that, however, it would be necessary to construct the
theory beyond manifolds with reflection.  The most direct way to do this would be to assign an element
of the Hilbert space $H(\Sigma)$ to any manifold with boundary $\Sigma.$  This in turn would arise
from a map $\cF_+ \to H(\Sigma)$ as discussed in Remark \ref{moralpf}.

Another question is the relation of our construction to the topological quantum field theory
of \cite{Witten,rt}. The Hilbert space we have
associated to a two-manifold here is infinite-dimensional, whereas
the constructions of \cite{Witten,rt}, arising from rational conformal field theory,
give finite dimensional vector spaces.  On the other hand, we have
built our construction for generic values of the coupling $\lambda$ lying on the real line
whereas the constructions of \cite{Witten,rt} require special values
of the coupling lying in a different region of the complex plane.  One analogy is to the representations of finite
dimensional lie algebras, where for any choice of highest weight,
one obtains an infinite
dimensional Verma module;  for dominant integral choices of this weight, this
infinite dimensional module has a finite dimensional quotient.  It is
plausible that some similar phenomenon might appear here, where the 
generic quantum field theory we have constructed may be ``analytically
continued'' (in analogy to the types of analytic continuation that
give Minkowskian Quantum Field Theory in terms of Euclidean Quantum
Field Theory) to values of the couplings where such a finite dimensional
quotient may arise.  In the context of knot theory, we may speculate that
the representations of the Braid group constructed by Jones 
\cite{Jones} may be related to
 quotients of sufficiently high tensor powers of the Burau representation.

\subsection{String Field Theory} In a more speculative vein, proposals for String Field Theory have also been presented which involve a cubic functional
integral not unlike that appearing in Chern-Simons gauge theory.  We may
also hope by analogy that ``finiteness in second order perturbation theory
implies existence of the nonperturbative theory in imaginary coupling,'' as for the $\phi^3$ theory described in Section \ref{ccqft}.  For
superstring theory, the finiteness of second order perturbation theory
has been established by d'Hoker and Phong \cite{dp}.  If these superstring 
theories can be written in a satisfactory form as String Field Theory,
this may hint at such a construction.

\vfill\eject
\appendix\section{Formulas for the Multiplication action}
 
In this section we briefly review a multiplication formula for the fields in the case where $M$ is split into three components by submanifolds isotopic to $\Sigma,$ which we will not use directly, but which we record to show the topological nature of the theory.   Suppose $\Sigma_1,\Sigma_2$ are two compact connected,  disjoint submanifolds of $M$, each isotopic to $\Sigma. $ Each of these splits  $M$ into two components, and  the union $\Sigma_1 \cup \Sigma_2$ splits $M$ into three components $U, V, W$ (see Figure 1).

 \begin{figure}\labell{figure2}
       \includegraphics[width=0.85\textwidth]{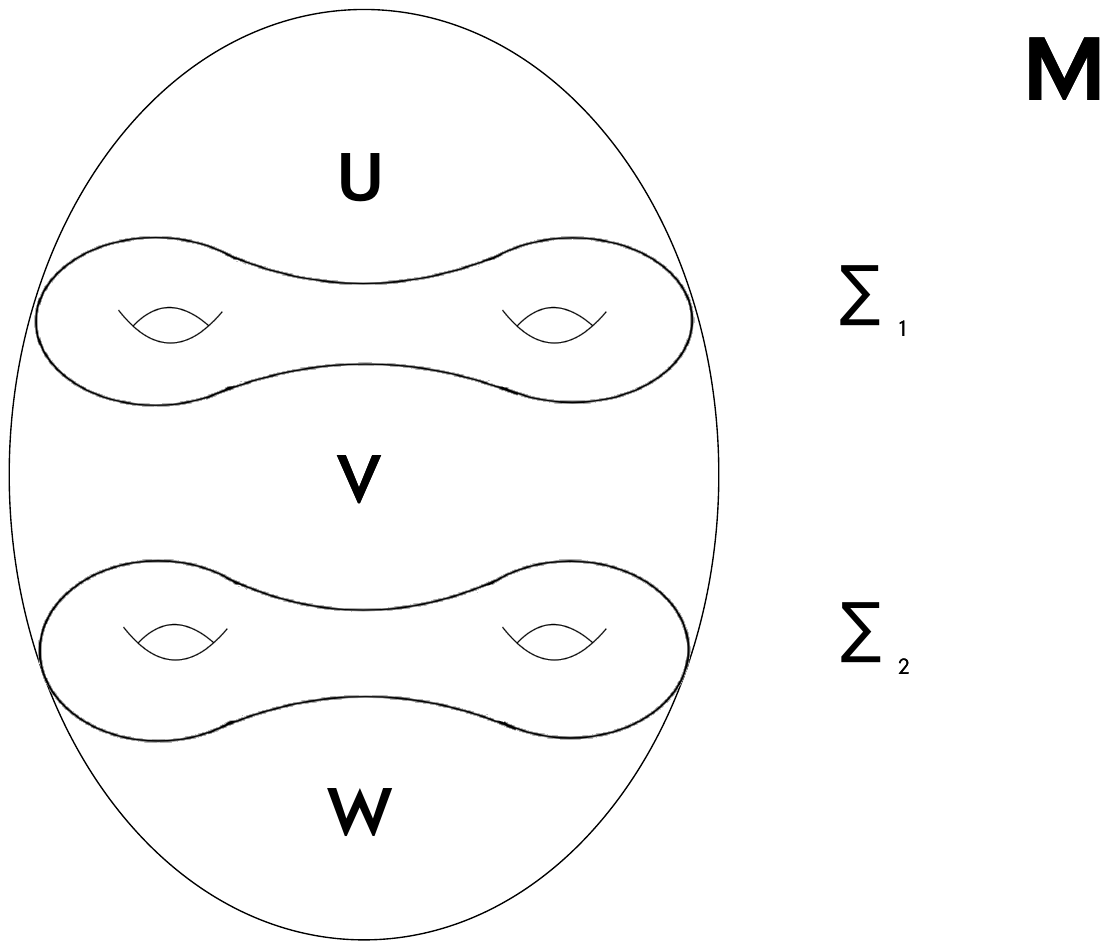}  
    \caption{The manifold $M$ cut into three open sets by two isotopic submanifolds $\Sigma_1,\Sigma_2$}
   \end{figure}

We begin with formulas for the gauge fields.  Using Proposition \ref{SplitM}, we have

\begin{Lemma}\labell{twosplittingb} 
Suppose $p \in \cF^B_P(U), q\in \cF^B_P(V), r \in \cF^B_P(W).$
 Then
  $$
\Phi(pqr) = \langle   \hat{\pi}^B_{\Sigma_1} (p),  S^*\hat{\pi}^B_{\Sigma_1} (q r)\rangle_{H^B(\Sigma_1) }=
\langle   \hat{\pi}^B_{\Sigma_2} (pq) ,    S^*\hat{\pi}^B_{\Sigma_2} ( r) \rangle_{H^B(\Sigma_2) };$$

\noindent where elements of ${\rm Sym}^*\Lambda \subset {\rm Sym}^*H^1( {\Sigma_1},\fg)$ and ${\rm Sym}^*\Lambda \subset{\rm Sym}^*H^1( {\Sigma_2},\fg)$ are identified with elements
of $H^B(\Sigma)$ using the isotopy.   
\noindent Also,

$$\hat{\pi}^B_{\Sigma_1}(p) = \hat{\pi}^B_{\Sigma_2}(p).{}\footnote{Note that as in Proposition \ref{splitM}, this indicates that the Hamlitonian in this theory is zero.}$$
 
\end{Lemma}  

Similarly for the ghosts, using Proposition \ref{piisom}, we have the following

\begin{Lemma}\labell{twosplitting} 
Suppose $p \in \cF^F_P(U), q\in \cF^F_P(V), r \in\cF^F_P(W).$
Then $pq \in \cF^F(U\cup V), qr \in \cF^F(V \cup W), $ and 
$$
\Psi(pqr) = \langle   \pi^F_{\Sigma_1} (p^c), \cJ_F \pi^F_{\Sigma_1} (q r)\rangle_{H^F(\Sigma_1) }=
\langle   \pi^F_{\Sigma_2} (q^cp^c) ,   \cJ _F \pi^F_{\Sigma_2} ( r) \rangle_{H^F(\Sigma_2) };$$
 
\noindent and

$$\pi^F_{\Sigma_1}(p) = \pi^F_{\Sigma_2}(p).$$

\end{Lemma} 

Combining Lemma \ref{twosplitting} with Lemma \ref{twosplittingb},  we then have the following formula for polynomials in the Bose and Fermi fields.  Let $\Sigma_1,\Sigma_2, U, V, W$ be as above.  Assume that the orientation of $\Sigma_1$ obtained from the isotopy agree with the orientation of the boundary of $U.$  Write

$$\hat{\pi}_{\Sigma_i} = \hat{\pi}_{\Sigma_i}^B \otimes {\pi}_{\Sigma_i}^F, i = 1,2.$$

Then\footnote{Note that the action of $S^*$ on the even cohomology of $\Sigma$ amounts to one negative sign on $H^2,$ which is morally accounted for by the signs in the adjoint $(\cdot)^c.$}

\begin{Proposition}\labell{twosplittingbf} 
Suppose $p \in \cF_P(U)  , q \in \cF_P(V) , r \in \cF_P(W)  .$
Then  
$$(\Phi \otimes \Psi)(pqr) = \langle   \hat{\pi}_\Sigma (p^c),  \cJ (S^*\otimes {\rm id}) \hat{\pi}_{\Sigma_1} (q r)\rangle_{H (\Sigma_1) }=
\langle   \hat{\pi}_{\Sigma_2} (q^cp^c) ,   \cJ (S^*\otimes {\rm id})\hat{\pi}_{\Sigma_2} ( r) \rangle_{H(\Sigma_2) },$$ 

\noindent where we extend the action of $S^*$ on $H^B(\Sigma)$ to the operator $S^*\otimes {\rm id}$ on $H(\Sigma)$ by defining its action on Fermi fields to be the identity.  Likewise we define the action of $(\cdot)^c$ to  $\cF_P(U) (U)$ by defining its action on Bose fields to be the identity.
\noindent Furthermore

$$ \hat{\pi}_{\Sigma_1}(p) =  \hat{\pi}_{\Sigma_2}(p).$$
 
\end{Proposition}

\end{document}